\documentclass[esd, manuscript]{copernicus}
\usepackage{url}

\begin{document}
\nolinenumbers

\title{The role of internal variability in global climate projections of extreme events}


\Author[1, *]{Mackenzie L.}{Blanusa}
\Author[2]{Carla J.}{L\'{o}pez-Zurita}
\Author[2, $\dag$]{Stephan}{Rasp}

\affil[1]{University of Connecticut, Avery Point Campus, 1084 Shennecossett Rd, Groton, CT, 06340}
\affil[*]{Work done during an internship at ClimateAi}
\affil[2]{ClimateAi, San Francisco}
\affil[$\dag$]{Now at Google Research}




\correspondence{Mackenzie Blanusa (mackenzie.blanusa@uconn.edu)}

\runningtitle{Internal variability of extremes}

\runningauthor{Blanusa et al.}

\received{}
\pubdiscuss{} 
\revised{}
\accepted{}
\published{}


\firstpage{1}

\maketitle

\begin{abstract}
Climate projection uncertainty can be partitioned into model uncertainty, scenario uncertainty and internal variability. Here, we investigate the different sources of uncertainty in the projected frequencies of daily maximum temperature and precipitation extremes, which are defined as events that exceed the 99.97th percentile. This is done globally using initial-condition large ensembles.
For maximum temperature extremes, internal variability dominates in the next two decades. Around the middle of the 21st century model and scenario uncertainty become the dominant contribution in the tropics but internal variability remains dominant in the extra-tropics. Towards the end of the century, model and scenario uncertainty increase to near equal contributions of $\sim$40\% each globally with large regional fluctuations.
For precipitation extremes, internal variability dominates throughout the 21st century, except for some tropical regions, for example, West Africa. 
In regions where internal variability constitutes the major source of uncertainty, the potential impact of reducing model uncertainty on the signal-to-noise ratio of the climate projection is estimated to be small. 
We discuss the caveats of the methodology used and impact of our findings for the design of future climate models. The importance of internal variability found here emphasizes that large ensembles are a vital tool for understanding climate projections. 
\end{abstract}


\introduction  

Uncertainties in climate projections arise from three different sources \citep{hawkins_potential_2009}. First, there is \textit{model uncertainty}. Climate models have errors in their representation of dynamical and physical processes and disagree in how they respond to external forcings, most notably to increases in greenhouse gases. Second is \textit{scenario uncertainty}, which describes the ambiguity related to how greenhouse gas emissions will develop in the future. Third and last, there is \textit{internal variability} which is caused by the chaotic nature intrinsic to the Earth system. To take a simple example, a decision maker might be interested in how many extreme heat days to expect in the coming decade. Even in a stationary climate, the outcome is uncertain, as, by chance, some decades see many heat extremes while others only have few. How much each of these three uncertainties contributes to the total uncertainty depends on the parameter of interest, the geographic location and how far ahead one is projecting. 

Naturally, for decision making one would like to reduce the uncertainty of climate projections. However, while one can hope to reduce model uncertainty by building better climate models, internal variability sets a hard limit on how low uncertainty can become, similar to the intrinsic limit of predictability in weather forecasting \citep{zhang_what_2019}. Therefore, if internal variability makes up a large fraction of the total variability, even a significant model improvement would only lead to a minor reduction in total uncertainty. Scenario uncertainty is a curious case. Often it is assumed to be irreducible similar to internal variability, especially as commonly used climate model scenarios do not have likelihoods attached to them. However, in some use cases decision makers look at a single scenario at a time. One example is mitigation, where some influence on future emissions is assumed.

To study the partitioning of variability requires separating the forced response from the chaotic fluctuations for a given climate model. This was first done by fitting a smooth curve to time series of annual temperature and precipitation and estimating internal variability from the residual by \cite{hawkins_potential_2009} and \cite{hawkins_potential_2011} (hereafter abbreviated HS09 and HS11, respectively). 
One issue with their fitting method is that it breaks down for noisy time-series, e.g. regional precipitation extremes \citep[hereafter referred to as L20]{lehner_partitioning_2020}. This is where large ensembles can help out, in which a single model is run multiple times with varying initial conditions \citep{deser_insights_2020}. Using large ensembles, one can estimate the models forced response as the ensemble mean and the internal variability from the ensemble variance. Because of their usefulness, more and more modeling centers are running large ensembles. A third way to estimate internal variability is by using an observational large ensemble. These are based on statistical models of different modes of variability (e.g. ENSO) with parameters fitted to the observational record. Because of this constraint to use past observations they generally are only used to provide several potential realizations of the past rather than future warming scenarios \citep{mckinnon_internal_2018, mckinnon_inherent_2021}. 

Regardless of the methodological details, studies largely agree on the role of internal variability. For annual mean temperature, internal variability plays a large role in short-term projections (up to 15 years), while the medium-term (30-40 years) is dominated by model uncertainty. Towards the end of the century, scenario uncertainty becomes the dominant factor in many regions of the earth, while internal variability becomes negligible (HS09, L20,  \cite{maher_quantifying_2020}). When looking at regional uncertainty, internal variability usually plays a larger role, often making up $>50\%$ of the total variability for most of the 21st century \citep{rasmussen_probability-weighted_2016}. For precipitation, a more chaotic variable, internal variability plays a larger role and can stay dominant even through the middle of the century regionally (HS11, L20). Further, since precipitation is less directly influenced by global warming, scenario uncertainty makes up a smaller fraction of the total uncertainty in most regions even at the end of the century. HS11 also introduced the use of a signal-to-noise ratio (SNR), dividing the expected change over the total uncertainty. For average temperature the SNR is usually quite large, while it is much smaller for precipitation. Looking at less averaged events or even extremes further increases the relative importance of internal variability. Examples of this are shown in several studies. L20 (their Fig.~7) showed that internal variability dominates for decadal mean winter precipitation in Seattle until the end of the century and for annual mean summer temperatures in Southern Europe until the middle of the century. \cite{deser_insights_2020} looked at the internal variability of extreme heat days in Dallas, Texas and found a large inter-ensemble spread (their Fig.~3). \cite{suarez-gutierrez_internal_2018} investigated the internal variability of European summer heat extremes at given 1.5 or 2$^{\circ}$C warming levels and showed significant overlap in the distribution of the two scenarios.

In this study, we aim to build upon prior research by systematically studying the role of internal variability for daily temperature and precipitation extremes on a global scale. Specifically, we want to answer the following questions:
\begin{enumerate}
    \item What fraction of the total uncertainty is made up by internal variability as a function of location and forecast time?
    \item What is the signal-to-noise ratio, a rough proxy for the actionability of a climate projection?
    \item How much can one hope to reduce total uncertainty by minimizing model uncertainty?
\end{enumerate}

\section{Methods}
\label{sec:methods}

\subsection{Data}
The large ensembles for this study were chosen because of their availability in the cloud. Climate projections, and large ensembles in particular, consist of huge amounts of data. Direct cloud access using the Zarr data format\footnote{\url{https://zarr.readthedocs.io/en/stable/}} eliminates a big data engineering hurdle, enabling studies such as this one for groups that do not already have the data downloaded on a local server. We acknowledge that more large ensembles exist, for example those included in Multi-Model Large Ensemble Archive (MMLEA) \citep{deser_insights_2020} and additional ones in CMIP6, but we did not use them as they were not available in the cloud. 

Our variables of interest are daily maximum temperature (\textdegree{C}) and daily precipitation (mm). For all models, daily data was accessed globally and regridded to a 2.5 x 2.5 degree grid (as in L20) using the bilinear interpolation in the xesmf Python package \footnote{\url{https://doi.org/10.5281/zenodo.1134365}}.

\subsubsection{Large ensembles}

Five large ensembles were used in this study and are listed in Table \ref{tab:Table.1}. All large ensembles except CESM-LE are sourced from the CMIP6 cloud archive on Pangeo's Google Cloud Storage bucket \footnote{\url{https://catalog.pangeo.io/browse/master/climate/cmip6_gcs/}}. CESM-LE, a CMIP5-class model, is hosted on the AWS cloud \footnote{\url{https://doi.org/10.26024/wt24-5j82}}. The number of ensemble members ranges from 10 to 58. Only models with 10 or more members were used, as models with fewer realizations would not provide adequate sample size. Even 10 members is small, as ensembles with less than 30 members have been found to underestimate variability \citep{wood_changes_2021}. We used daily data from historical (up to 2014) and SSP5-8.5 (2015-2100) simulations for all CMIP6 large ensembles. For the CESM-LE large ensemble, we used the 20C and RCP8.5 simulations. There are differences between the SSP5-8.5 and RCP8.5 scenarios but in the context of this study, we deemed them negligible. Because CESM-LE is a CMIP5-class model, the historical period only extends to 2004 while the RCP8.5 scenario starts in 2005. To make these simulations compatible with the CMIP6 models, we concatenated the two simulation periods and subsequently split them in 2015. This should not matter given our definition of extremes using quantiles, see below.

\begin{table}[ht]
\caption{List of large ensembles used in this study}
\label{tab:Table.1}
\begin{tabular}{|l|l|l|l|l|l|l|}
\hline
\textbf{Modeling Center} & \textbf{Model Version} & \textbf{Resolution (atm/oc)} & \textbf{Years} & \begin{tabular}[c]{@{}l@{}}\textbf{No. of members} \\ \textbf{(historical, future)}\end{tabular} & \textbf{Forcing} & \textbf{Reference} \\ \hline \hline
MIROC & MIROC6 & $\sim$1.4° x 1.4° / 1° & 1850-2100 & 50, 50 & \begin{tabular}[c]{@{}l@{}}historical, \\ ssp585\end{tabular} & \citet{tatebe_description_2019} \\ \hline
NCAR & CESM-LE & $\sim$1.3° x 0.9° / 1° & 1920-2100 & 40, 40 & \begin{tabular}[c]{@{}l@{}}20C, \\ rcp8.5\end{tabular} & \citet{kay_community_2015} \\ \hline
CCCma & CanESM5 & $\sim$2.8° x 2.8° / 1° & 1850-2180 & 50, 50 & \begin{tabular}[c]{@{}l@{}}historical, \\ ssp585\end{tabular} & \citet{swart_canadian_2019} \\ \hline
MPI & MPI-ESM1-2-LR & $\sim$1.9° x 1.9° / 1.5° & 1850-2100 & 10, 10 & \begin{tabular}[c]{@{}l@{}}historical, \\ ssp585\end{tabular} & \begin{tabular}[c]{@{}l@{}}\citet{muller_higher-resolution_2018}, \\ \citet{mauritsen_developments_2019}\end{tabular} \\ \hline
EC-Earth-Consortium & EC-Earth3 & $\sim$0.7° x 0.7° / 1° & 1850-2100 & 71, 58 & \begin{tabular}[c]{@{}l@{}}historical, \\ ssp585\end{tabular} & \citet{doscher_ec-earth3_2022} \\ \hline
\end{tabular}
\end{table}

Looking at the results, we noticed that the CanESM5 model consistently showed much larger trends in most regions compared to the other models (see Fig.~\ref{fig:ensemble line plot with CanESM5}, Seattle, as an example). As it turns out, CanESM5 is the most extreme example of the "hot model" problem\footnote{\url{https://www.carbonbrief.org/guest-post-how-climate-scientists-should-handle-hot-models/}} in CMIP6 (see Supplement of \cite{zelinka_causes_2020}), where cloud feedbacks lead to---what are generally considered---unrealistically high climate sensitivities in some models. Including CanESM5 leads to significanly larger values of model variability (Fig.~S12). In the end, we decided to remove CanESM5 from our variability computations since its climate sensitivity is far outside the range considered plausible. A similar process is done in the latest IPCC report \citep{hausfather_climate_2022}. The four remaining models represent a somewhat evenly spaced sampling of climate sensitivities according to \cite{zelinka_causes_2020}.

We also found weird behavior of the MPI-ESM1-2-LR model in West Africa, where temperature variability drops drastically in the future projections compared to the historical simulations, alongside a drop in mean temperature, and precipitation variability increases (see Fig.~\ref{fig:ensemble line plot with CanESM5}, Lagos). We do not know the cause for this. In this instance, we decided to include the MPI-ESM1-2-LR model in our subsequent analysis for two reasons. First, the suspicious behavior was restricted regionally; second, even when removing the model, the variability results do not change drastically because other models also contribute to the spread in West Africa. It is important to keep in mind though that model variability could be erroneously inflated in West Africa.

\subsubsection{CMIP6 single realizations}

Most large ensembles are only run for a single, high-emission scenario.
To obtain an estimate of scenario uncertainty, we follow L20 and use single realizations of CMIP6 models for four scenarios: SSP1-2.6, SSP2-4.5, SSP3-7.0, and SSP5-8.5. Fourteen CMIP6 models were used in this study, because of their cloud availability of daily data for all scenarios and variables required. These models are CMCC-CM2-SR5, CMCC-ESM2, EC-Earth3, EC-Earth3-Veg-LR, GFDL-ESM4, IITM-ESM, INM-CM4-8, INM-CM5-0, IPSL-CM6A-LR, KACE-1-0-G, MIROC6, MPI-ESM1-2-HR, MPI-ESM1-2-LR, and NorESM2-MM. 

\subsection{Definition of extremes}

Every end user has a different definition of what an extreme is. Here, we use a quantile-based definition but acknowledge that this is not a one-size-fits-all solution. 
Specifically, we choose the 0.9997th quantile computed over a reference period, here chosen to be 1995-2014. This corresponds to the largest daily value in a 10 year period. Note that this is different from the usual definition for return periods which uses the yearly maximum rather than daily values. We still use the "1-in-10-year" terminology for convenience in some places in this paper. A comparison to using yearly maximum return periods can be found in the Supplement (Sec.~S4). Further, we investigated the sensitivity of the key results to the quantile chosen (Fig.~S1). Looking at larger quantiles (e.g. corresponding to a "1-in-100-year" event) generally leads to somewhat larger relative contribution of internal variability but the differences are, in most cases, within 20\% (as measured by the contribution of model to total uncertainty). We also tested using multi-day averages rather than daily values to account for longer-lasting extremes but found negligible effects ($<10\%$ with rare exceptions) on the results (Fig.~S2).

The quantile values are computed for each model separately, thereby performing an implicit quantile bias correction (see Sections~\ref{sec:discussion} and ~S2). The projections are then thresholded to yield a boolean time series with 1's indicating where an extreme event occurs. Next, we aggregate the data by year. Note that consecutive days where the extreme threshold is crossed are counted as separate events. Coarsening the data to avoid this has a negligible effect on the results (Fig.~S3). Subsequently, we average the yearly sums over an aggregation period, another hyper-parameter, yielding the average yearly number of extreme events in that period. As expected, this hyper-parameter has a large impact on the results (Fig.~S4). Larger aggregation periods lead to much smoother results and therefore less internal variability. To choose a reasonable value, we take inspiration from our operational work but also many previous studies, as well as the IPCC report \citep{arias2021climate} and climate disclosure frameworks. There, the impact of climate change is often considered on decadal time scales, which is why we chose 10 years as an aggregation period. 

The final time series $x(m,e,t,l)$ then contains the yearly average number of extreme events in each ensemble member in a 10 year window around the given year and is a function of model $m$, ensemble member $e$ (except for the single-realization CMIP6 models), time $t$ and location $l$.

\begin{figure}[hb!]
\includegraphics[width=\linewidth]{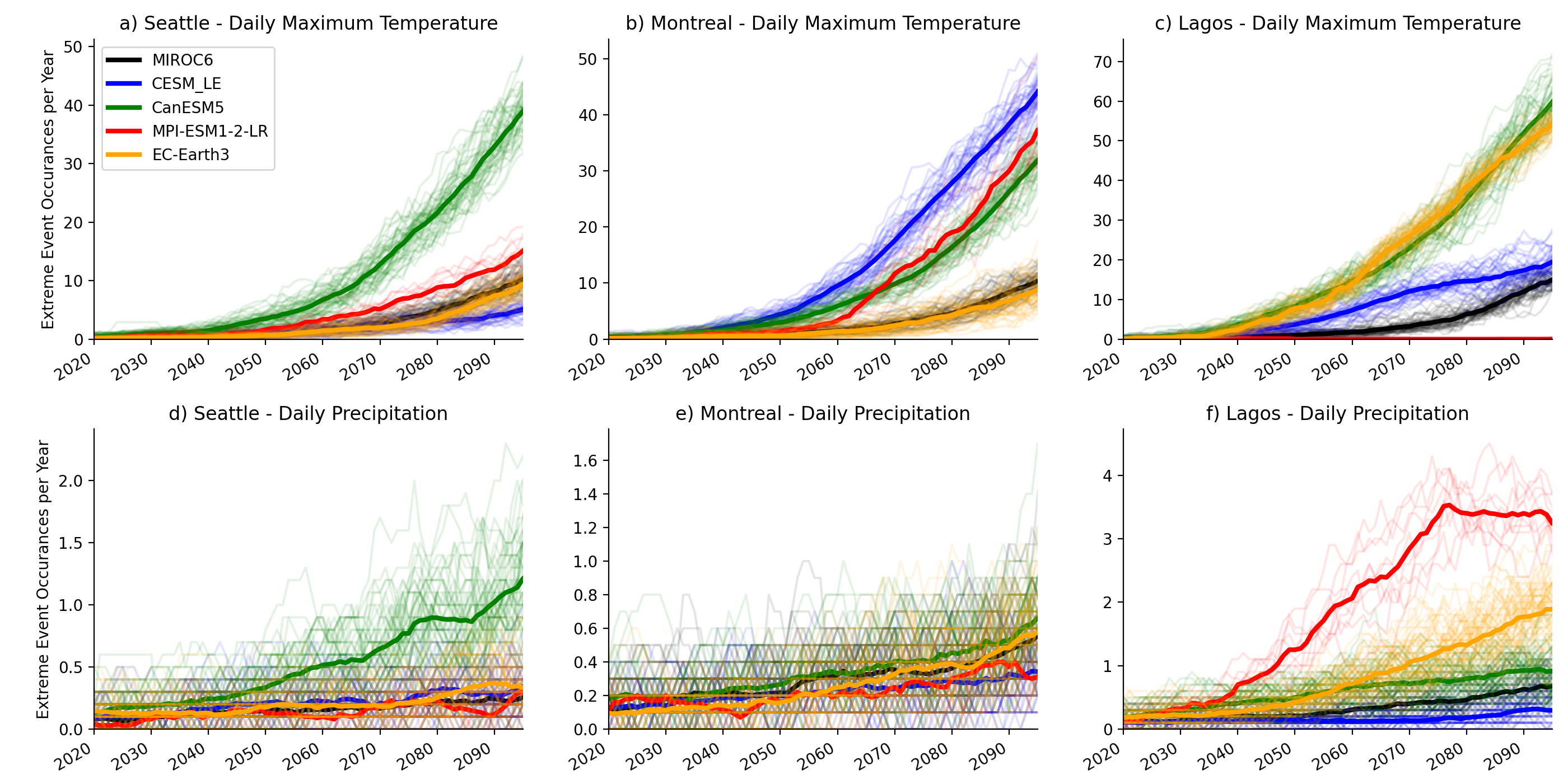}
\caption{Extreme event occurrence (days) per year from 2020-2095. Top row: daily maximum temperature for three locations; (a) Seattle, U.S.A., (b) Montreal, Canada, (c) Lagos, Nigeria. Bottom row: daily precipitation for the same three regions (d-f). Each data point represent the average yearly occurrence for a 10 year period around that year.}
\label{fig:ensemble line plot with CanESM5}
\end{figure}

\subsection{Partitioning of variability}
Total uncertainty $T$ is the sum of internal variability $I$, model uncertainty $M$ and scenario uncertainty $S$:
\begin{equation}
\label{eqn:total uncertainty}
    T(t,l) = I(t,l) + M(t,l) + S(t,l)
\end{equation}

We follow the methodology of L20 to compute the terms above from large ensembles. We also compare this to the fitting method of HS09 in the Supplement.

Internal variability for each model is computed as
\begin{equation}
\label{eqn: internal LE}
    I(m,t,l) = \text{var}_{e}(x)
\end{equation}
where $\text{var}_{e}(x)$ is the variance computed in the model dimension.
We can then average in the model dimension, yielding an average estimate of internal variability:
\begin{equation}
\label{eqn:internal average}
   I(t,l) = \text{mean}_{m}(I(m,t,l))
\end{equation}

Model uncertainty can be computed as the variance across ensemble means of each model:
\begin{equation}
\label{eqn:model uncertainty LE}
    M(t,l) = \text{var}_{m}(\text{mean}_{e}(x))
\end{equation}

To compute scenario uncertainty, we use the 14 single-realization CMIP6 models. Following methods described in HS09 and L20, scenario uncertainty is calculated as the variance across multi-model means for each of the four scenarios, $s$:
\begin{equation}
\label{eqn:scenario uncertainty}
    S(t,l) = \text{var}_{s}(\text{mean}_{m}(x))
\end{equation}


\subsection{Ideal Total Uncertainty}
Out of the three uncertainties, model, internal, and scenario, only model uncertainty is generally assumed to be reducible. Internal variability could potentially be reduced in the first decade by initialization but we did not consider this further (HS09). We can therefore define an ideal total uncertainty, $T_{\text{ideal}}$, for a climate projection which assumes a model uncertainty of zero.
\begin{equation}
\label{eqn:ideal total uncertainty}
    T_{\text{ideal}}(t,l) = I(t,l) + S(t,l) 
\end{equation}
If $S$ is ignored, the ideal total uncertainty is simply internal variability.

By comparing total and ideal total uncertainty we can understand the potential for improving climate projections, which will vary in space and time. We can estimate a max percent improvement using total (\ref{eqn:total uncertainty}) and ideal total uncertainty (\ref{eqn:ideal total uncertainty}) estimates:
\begin{equation}
\label{eqn:max improvement}
    \text{Max \% Improvement} = (1 - \sqrt{T_{\text{ideal}}} / \sqrt{T}) \times 100
\end{equation}
Here, it is important to compute the improvement in standard deviation, or real unit, space rather than variance, or squared unit, space, since ultimately decision makers are interested in real units.

\subsection{Signal to Noise Ratio}
The signal to noise ratio (SNR) can be taken as a very rough proxy for the actionability of a climate projection. It was first used in this context by HS11. The signal is defined as the model-mean change in $x$ with respect to the reference period 1995-2014, where $x_{\text{ref}}(l) = \text{mean}_{m, e, t_{\text{ref}}}(x)$: 
\begin{equation}
\label{eqn:signal}
    \text{Signal}(t,l) = \text{mean}_{m, e}(x) - x_{\text{ref}}(l)
\end{equation}

The noise is simply the total uncertainty, as in Eq.~(\ref{eqn:total uncertainty}). Therefore, as the name suggests, SNR is the ratio of signal to the total uncertainty:
\begin{equation}
\label{eqn:SNR}
    \text{SNR}(t,l) = \text{Signal}(t,l) / \sqrt{T(t,l)}
\end{equation}
We also compute an ideal SNR using $T_{\text{ideal}}$. Low SNRs imply that the noise dominates the signal and the projections are likely not that informative. On the other hand, large SNRs indicate more certain predictions.

\section{Results}

\subsection{Example locations}

\begin{figure}[ht]
\includegraphics[width=\linewidth]{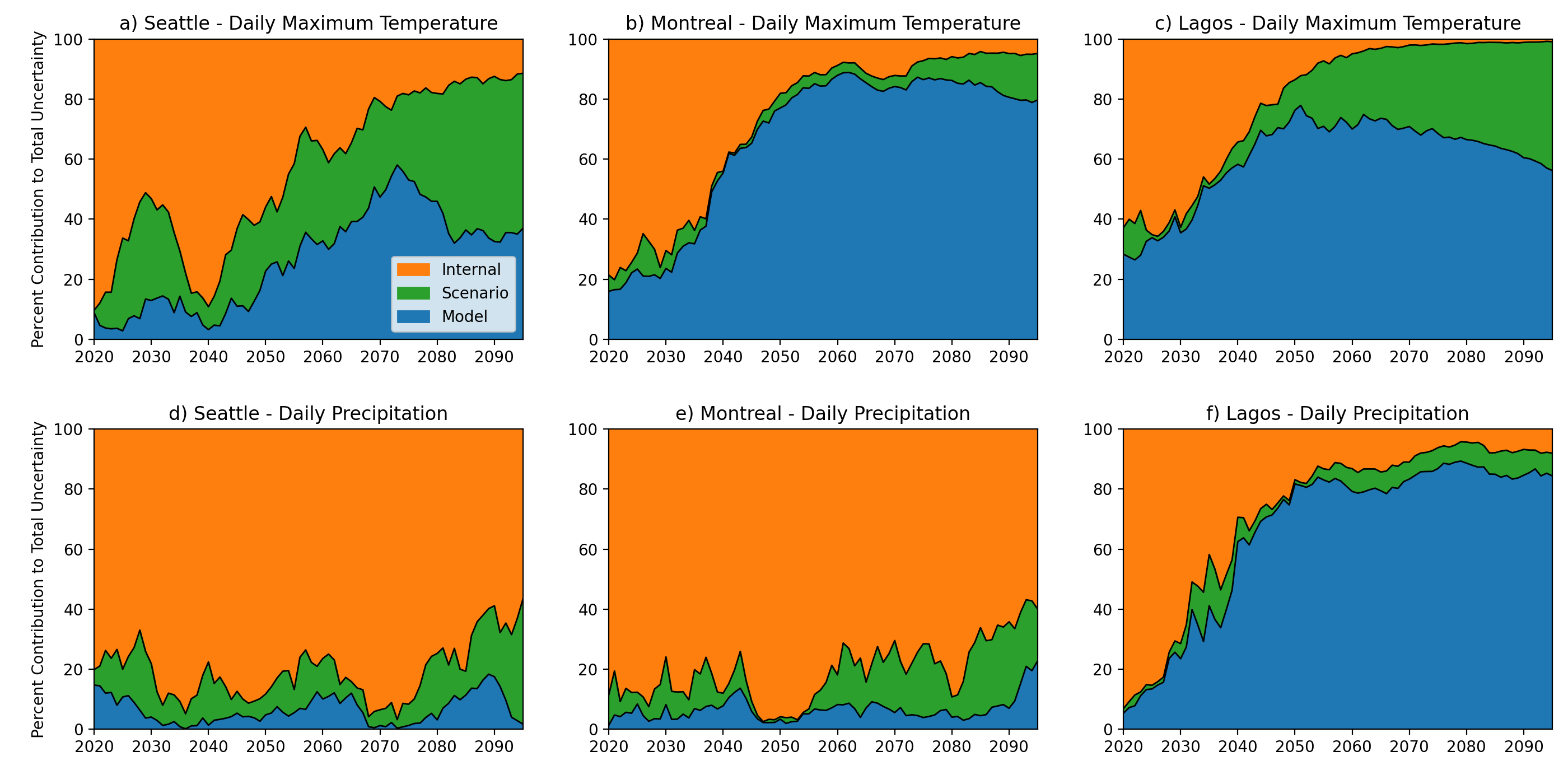}
\caption{Percent contribution to total uncertainty for model uncertainty, scenario uncertainty, and internal variability from 2020-2095 for three locations: Seattle, U.S.A. (a,d), Montreal, Canada (b,e), Lagos, Nigeria (c,f). Top row (a-c) is for daily maximum temperature and bottom row (d-f) for daily precipitation. }
\label{fig:percent contribution}
\end{figure}

We start by investigating variability at three locations: Seattle, USA, Montreal, Canada and Lagos, Nigeria (see top left panel of Fig. \ref{fig:global_tasmax} for map of locations). These locations were chosen in hindsight after seeing the global results because they represent different kinds of behavior. First, we look at the evolution of extreme occurrence, $x$, in the different ensembles (Fig. \ref{fig:ensemble line plot with CanESM5}). As per definition, each ensemble should start out with a value of around 0.1 occurrences per year in 2020. For maximum temperature, values tend to increase significantly towards the end of the century. In Seattle, the models roughly agree on how much extremes are expected to increase (ignoring CanESM5). In Montreal and Lagos, on the other hand, there is large disagreement between models, with one set of models projecting a roughly 5 times larger increase than another set. Projected extreme precipitation occurrence shows a much smaller signal compared to temperature, especially in Seattle and Lagos. The already mentioned behaviors of CanESM5 and MPI-ESM1-2-LR are clearly visible in these figures (see discussion above).

\begin{figure}[ht]
\includegraphics[width=\linewidth]{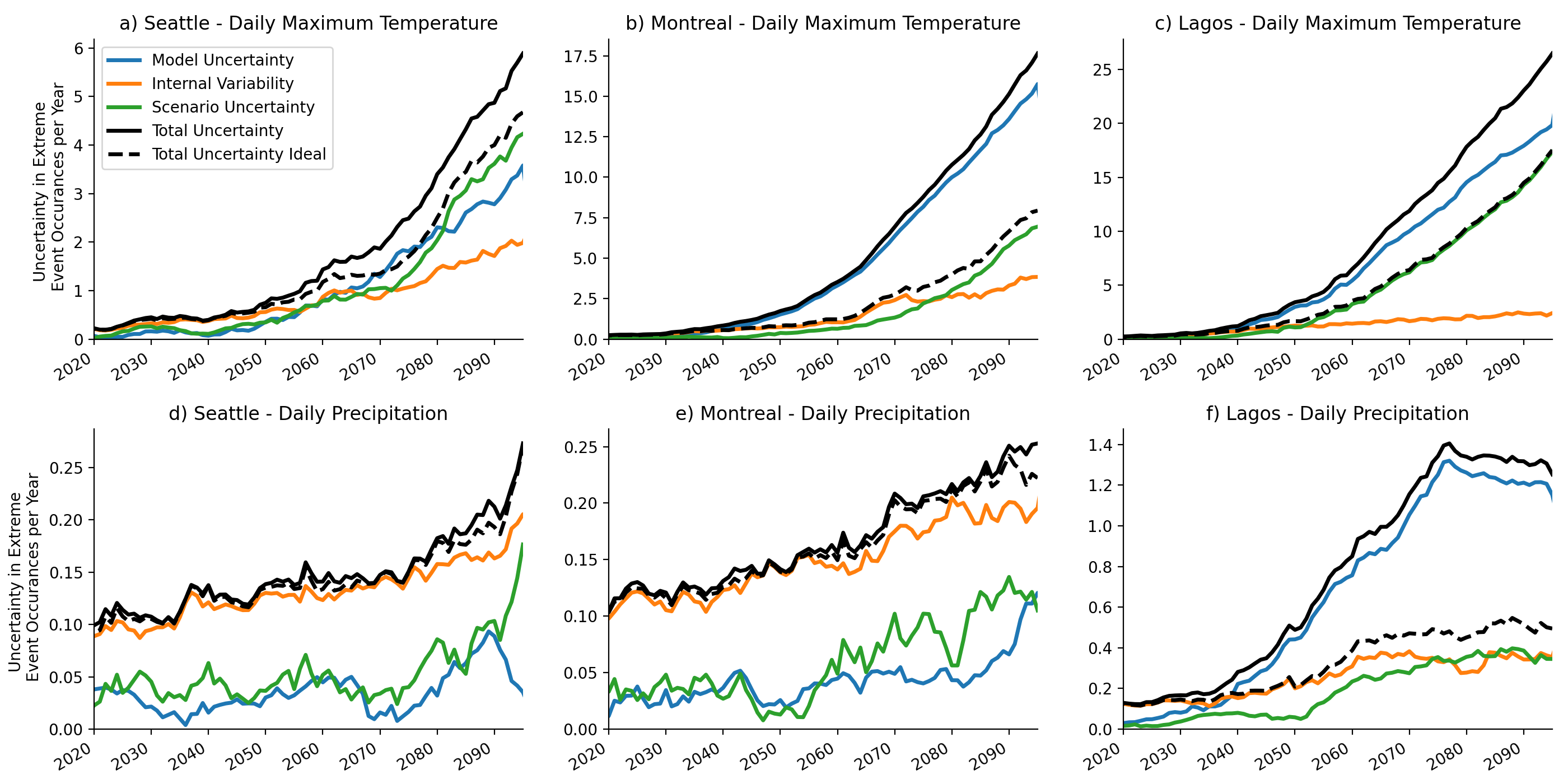}
\caption{Estimates of model uncertainty (blue), internal variability (orange), scenario uncertainty (green), total uncertainty (black, solid), and ideal total uncertainty (black, dashed, i.e. model uncertainty = 0) from 2020-2095 for three locations: Seattle, U.S.A. (a,d), Montreal, Canada (b,e), Lagos, Nigeria (c,f). Top row (a-c) is for daily maximum temperature and bottom row (d-f) for daily precipitation. Each data point represent the average yearly uncertainty (days) for a 10 year period around that year.}
\label{fig:all uncertainties}
\end{figure}

To show the different uncertainties, we show the classic partitioning plots following HS09 and L20 (Fig.~\ref{fig:percent contribution}). In addition, we also show the absolute values of the uncertainties as standard deviations (Fig.~\ref{fig:all uncertainties}). This is important since decision makers will be interested in the uncertainty in real physical units rather than squared units. The first thing that stands out is that the uncertainty estimates are relatively noisy, especially compared to previous literature. This is a result of the small sample size of our large ensembles but also of the noise inherent to extremes. We decided against further smoothing the plots to not obscure the noise in the method but note that the small-scale fluctuations are likely caused by small sample sizes and inaccuracies in the method (see Sec.~\ref{sec:discussion}).
One example is the uptick of $S$ for maximum temperature in Seattle around 2030. From looking at other locations and previous studies, one would expect the contributions to change relatively smoothly.

For maximum temperature, we see that for Seattle internal variability $I$ dominates with >50\% until the second half of the century, followed by an increase in model $M$ and, towards the end of the century, scenario variability $S$. In Montreal and Lagos we see a different picture. For the first century $I$ and $M$ are roughly equally important but model variability quickly makes up the largest contribution. $S$ increases towards the end, more so in Lagos, but $M$ remains the dominant contribution. Looking at the absolute variabilities in standard deviation space (Fig. \ref{fig:all uncertainties}) we immediately see that $I$ is not constant but rather increases significantly as total uncertainty increases. This is in contrast to previous studies of internal variability in mean quantities (e.g. HS09, HS11 and L20), where $I$ stays roughly constant in time. This also means that the fitting method of HS09 would be inappropriate for this context since it only computes a time-constant value for $I$. See Supplement for further comparison. 

For precipitation, in Seattle and Montreal, $I$ represents by far the largest contribution, throughout the century. Lagos, however, shows a large contribution of $M$ towards the end of the century, driven largely by the large projected increase in precipitation in the MPI-ESM1-2-LR model but also partly in the EC-Earth3 model. 

Fig \ref{fig:all uncertainties} also shows the total uncertainty and the ideal total uncertainty. One important thing to note is that the uncertainties only add up in variance space but not in standard deviation space. A simple example would be a case where $I = M = 1$ (ignoring $S$ for a second). In variance space the total uncertainty would be 2 but only $\sqrt{2} \approx 1.41$ in standard deviation space. One consequence of this is that in cases where $M$ makes up around 50\% of total uncertainty in variance space (see end of century temperature variability in Seattle and Lagos), setting $M = 0$, only reduces total uncertainty by around 30\%. In Montreal, where model uncertainty dominates, reducing $M$ has a much larger effect. Even here, however, where model uncertainty represents 80\% of total variance at the end of the century, the ideal total uncertainty in real unit space is still roughly 40\% of the actual total uncertainty.

\begin{figure}[ht]
\includegraphics[width=\linewidth]{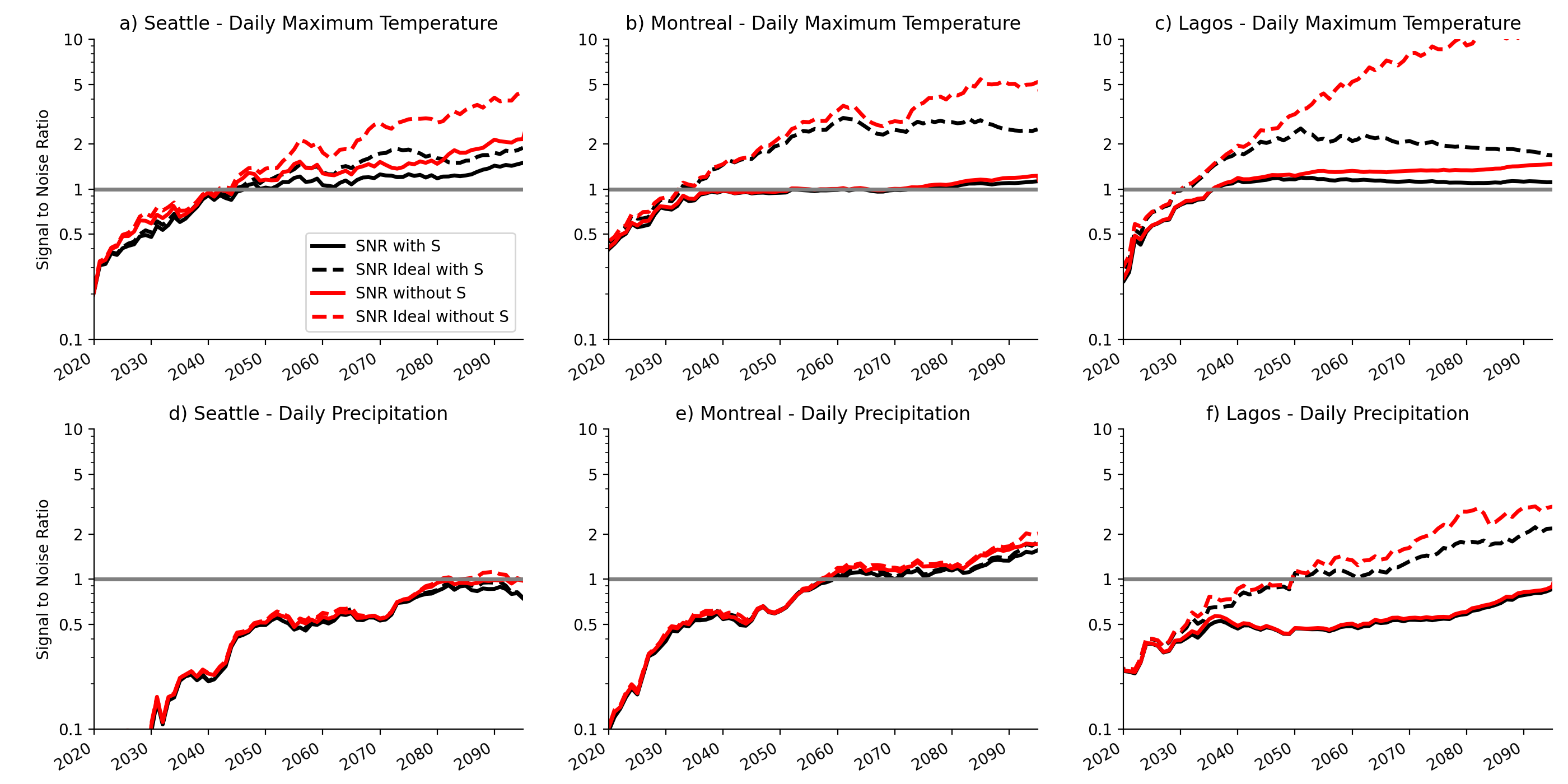}
\caption{Estimates of signal to noise ratio (SNR) from 2020-2095. Each panel includes SNR calculated with scenario uncertainty (black, solid), SNR ideal with scenario uncertainty (black, dashed), SNR without scenario uncertainty (red, solid), and SNR ideal without scenario uncertainty (red, dashed). Note the log-scale on the y-axis.}
\label{fig:SNR}
\end{figure}

Fig.~\ref{fig:SNR} shows the signal-to-noise ratio for four different cases. In black, the SNR including $S$ with the real $M$ in solid and the ideal case $M = 0$ in dashed. In red, the same but ignoring $S$. As mentioned above, SNRs significantly below one represent projections that are dominated by noise and, thus, most likely not very actionable. For maximum temperature, we can see that in Seattle the transition point SNR $=1$ is reached around 2040, regardless of the exact SNR case. Starting around 2050, one can see that decreasing $M$ could have a significant effect, especially when $S$ is ignored. For Montreal and Lagos, $M$ plays a much larger role. Model uncertainty included, SNR approaches one around 2030-2040 but then stays relatively constant around values of one. The ideal case of $M=0$ shows much larger SNRs from the start with values quickly going above 2 at around 2040. In Lagos, there is also a non-negligible impact of $S$, which becomes large towards the end of the century, thereby pushing the SNR down again. 

For precipitation, SNR values are very low at the start of the century in Seattle and Montreal, crossing SNR$=1$ right at the end of the century (Seattle) and in the 2050s (Montreal). Since $I$ is the dominant contribution, the four SNR cases are pretty much identical. In Lagos, once again, $M$ plays a big role. Including $M$ in the total variability leads to very low SNR values. Setting $M=0$, SNR reaches the transition point around the 2040, steadily increasing afterwards. 

\subsection{Global}

\begin{figure}[ht]
\includegraphics[width=\linewidth]{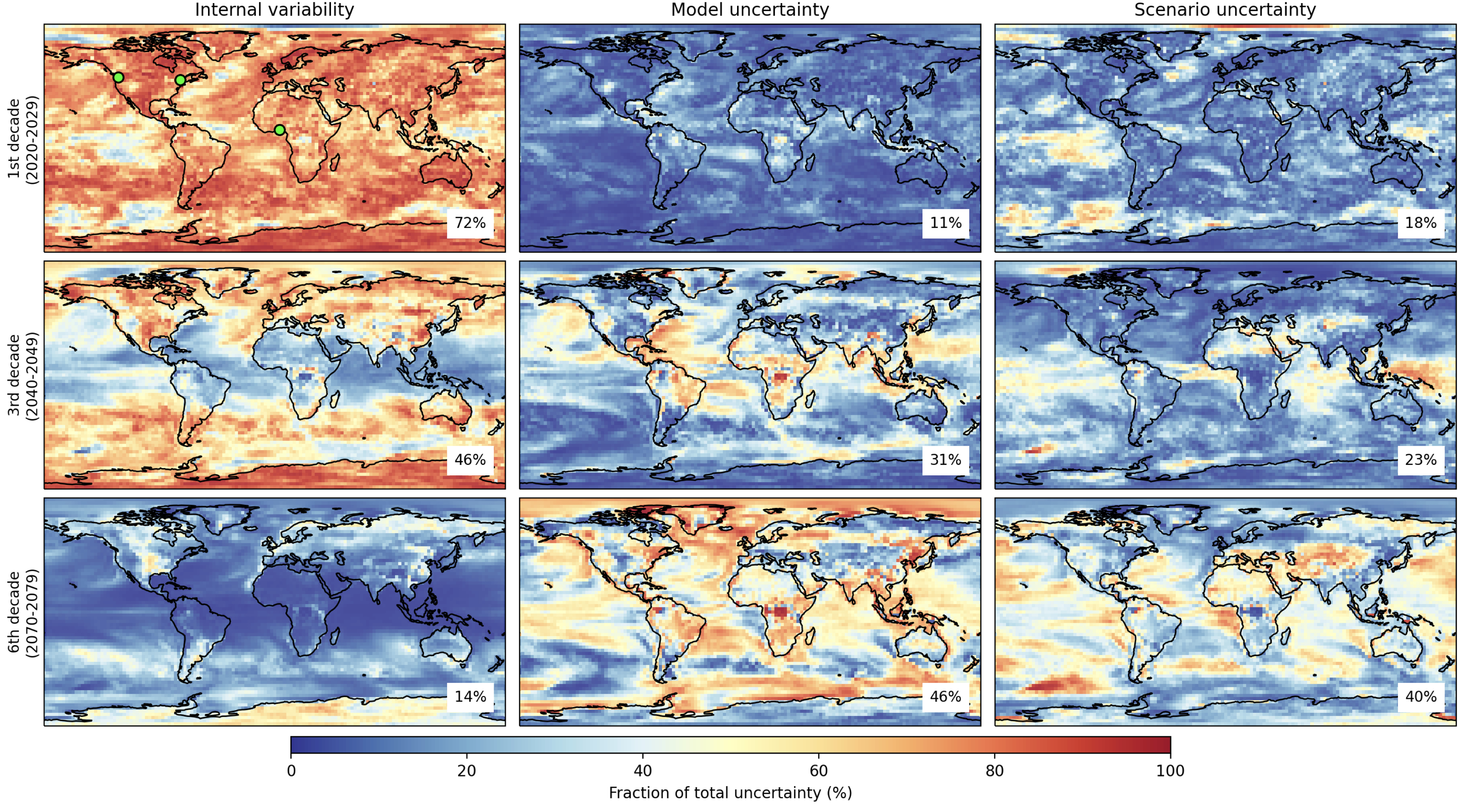}
\caption{Variance fractions of maximum temperature extremes for the three sources of uncertainty averaged for three decades, 2020-2029, 2040-2049 and 2070-2079. Percentage numbers are area-weighted global averages for each map. The three green dots in the top left panel mark three exemplary locations: Seattle, U.S.A., Montreal, Canada and Lagos, Nigeria.}
\label{fig:global_tasmax}
\end{figure}

\begin{figure}[ht]
\includegraphics[width=\linewidth]{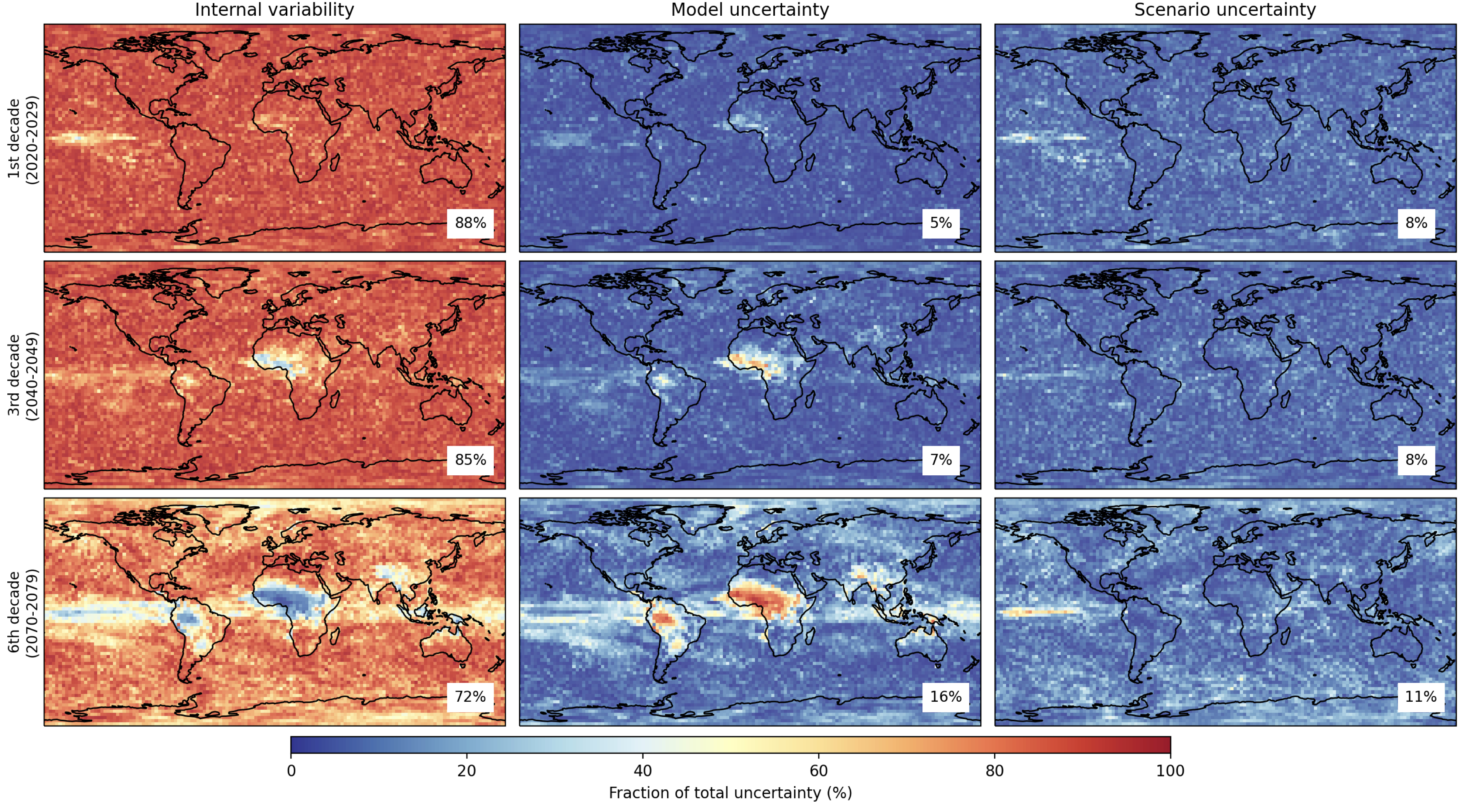}
\caption{As Fig.~\ref{fig:global_tasmax} but for precipitation extremes.}
\label{fig:global_pr}
\end{figure}

After looking at three exemplary locations in detail, we now generalize our diagnostics globally. First, similar to L20, we plot the fractional contributions of total variance for three decades (Figs.~\ref{fig:global_tasmax} and \ref{fig:global_pr}). Note that we chose earlier decades compared to L20. This is motivated by a focus on actionability and adaptation, where in our professional experience end users tend to be more interested in the next few decades rather than the end of the century. However, the results change smoothly in time and the qualitative difference between the time periods is small. 

For maximum temperature, $I$ is the dominant factor for most of the globe in the first decade. $M$ tends to be larger in the ITCZ area, the tropical continents and the storm tracks off the US East coast. There are some patches where $S$ is the largest uncertainty, for example in the East Pacific. Note, however that the absolute uncertainty and signal in the first decade is relatively small still. In the third decade, a rough division can be made between the tropics where $M$ and $S$ are important and the extra-tropics where $I$ remains dominant. There are however, large regional differences. The Atlantic off the US East coast and the Gulf of Mexico have large values of model uncertainty. Larger values of $I$ are also generally found towards the Eastern side of the larger continents with some exceptions for example the North Eastern US and Eastern Canada, as our example location Montreal shows. In the 2070s, $I$ plays a minor role except in some continental regions, for example the central US. $S$ and $M$ dominate but with a complex pattern between them that, at least somewhat, suggests some noise in the methodology. The results are largely in agreement with the decadal mean temperature variabilities in L20, even though some of the regional details differ. 

For precipitation the results are a lot more clear cut. $I$ is by far the largest contributor in most regions throughout the century. The only exception is West Africa, and to a lesser extent, the Amazon and the ITCZ where $M$ plays a larger role. As already explained, part of this behavior might be caused by some irregularities with the MPI-ESM1-2-LR model but even without this model the general pattern persists. This is also anecdotally confirmed by the Sahel decadal JJA rainfall results in L20. Partly, this hints at a "real" physical, underlying cause. However, it is also well known that coarse climate models struggle to represent the large convective systems present in these regions \citep{moncrieff_simulation_2017}. Further, as L20 speculate, this could also be caused by sampling different phases of the Atlantic Ocean's decadal variability \citep{yeager_predicting_2018}, which for decadal prediction is potentially fixable using observation-based initialization.

\begin{figure}[ht]
\includegraphics[width=\linewidth]{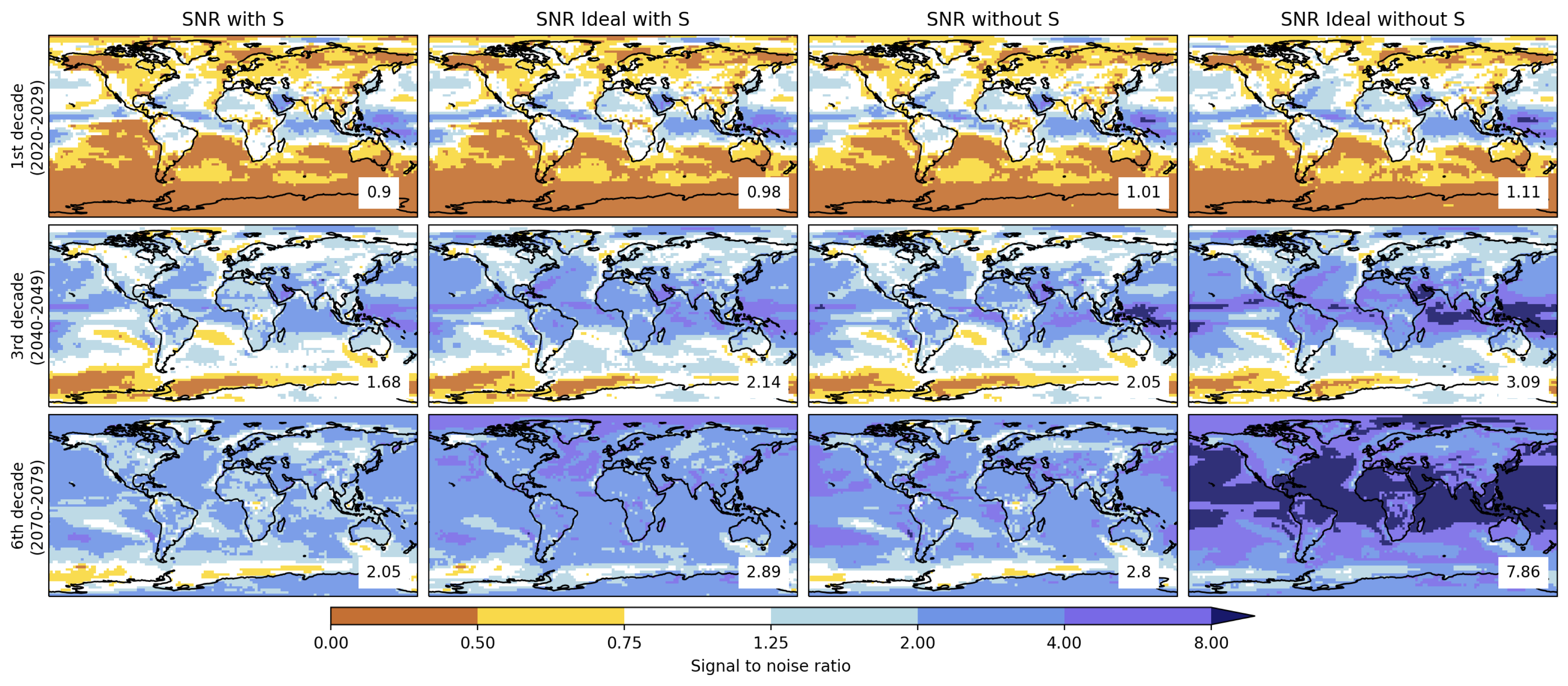}
\caption{Decade-average signal to noise ratio (SNR) for daily maximum temperature for four cases as described in the text.}
\label{fig:global_snr_tasmax}
\end{figure}

\begin{figure}[h]
\includegraphics[width=\linewidth]{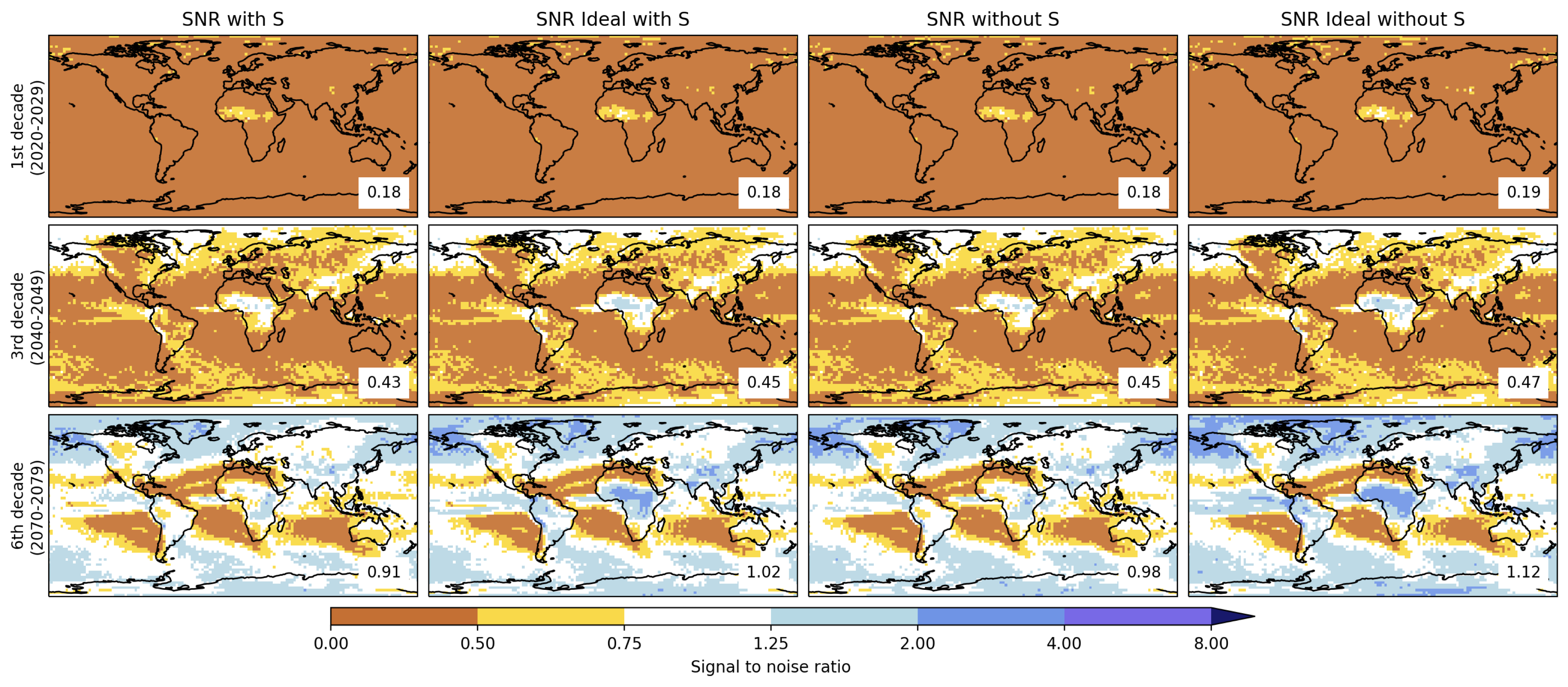}
\caption{As Fig.~\ref{fig:global_snr_tasmax} but for precipitation.}
\label{fig:global_snr_pr}
\end{figure}

Next we look at the four flavors of SNR, already discussed for the three example locations (Figs.~\ref{fig:global_snr_tasmax} and \ref{fig:global_snr_pr}). For maximum temperature, in the 1st decade SNR values are generally below one in the extratropical continents with little effect of setting $M$ or $S$ to zero since $I$ makes up the largest contribution by far. Some of the tropical continents have values of SNR slightly above one, with only the tropical maritime regions showing large SNR values, caused by the fact that variability there is generally very low. In the 2040s, SNR values in the continents tend to be slightly above one. Here the SNR rises noticeable in the "ideal" scenario of no model variability, especially over Africa and South America. In the 2070s, this trend is even stronger, with increases from setting $M=0$ seen globally to values above 2. The increase is even stronger if one considers the case without scenario uncertainty. 

For precipitation, we see low SNR values globally. Only in the latter half of the century do some regions see SNR $> 1$, for example eastern North America, Alaska, Greenland, eastern Eurasia and Central Africa. The latter region is also the only one to see a jump in SNR from setting $M=0$, a logical consequence of the large model variability discussed above. 

\begin{figure}[ht]
\includegraphics[width=\linewidth]{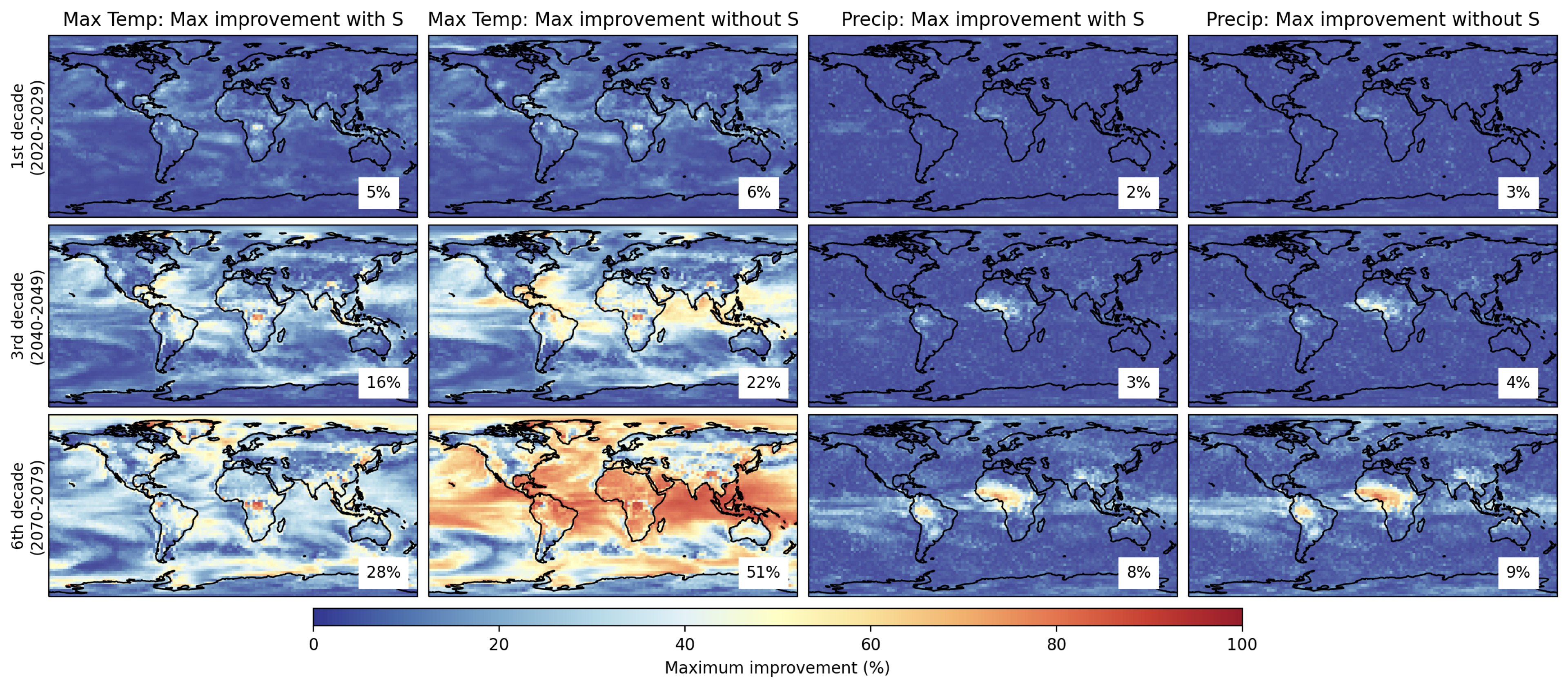}
\caption{Decade-average maximum \% improvement for maximum temperature and precipitation, including and excluding $S$.}
\label{fig:gobal_max_improv}
\end{figure}

As a final metric we look at the maximum possible improvement in total uncertainty by reducing model uncertainty to zero (Fig.~\ref{fig:gobal_max_improv}). For maximum temperature, values over the extra-tropical continents remain low ($< 30 \%$) for the first half of the century. Only some tropical regions show larger values. Even in the latter half of the century, most regions show a small maximum improvement if $S$ is considered. If $S$ is ignored, potential improvements jump and are quite high ($> 60\%$), even for some extra-tropical continental regions. As to be expected by now, the potential improvements for precipitation are very small throughout the century, except in the aforementioned tropical regions.

\section{Discussion}
\label{sec:discussion}

\subsection{Uncertainties in method}
For several reasons, which will be discussed below, the results presented in this paper, and that of similar studies, should be seen as a "ballpark estimate" rather than an exact assessment.

\subsubsection{Small sample size}
First, sample sizes are small for nearly all variance estimates. The smallest ensemble size is 10 members which, as seen in Fig.~\ref{fig:ensemble line plot with CanESM5}, results in a slightly noisier ensemble mean for the MPI-ESM1-2-LR ensemble compared to the other models, thereby picking up some internal variability as signal, similar to the fit method of HS09. \cite{wood_changes_2021} found that using small ensembles (less than 30 members) also results in an underestimation of variability. Next, the number of models is low. After removing CanESM5 because of its unrealistically high climate sensitivity only 4 models remain. Some confidence can be gained by the fact that using the fitting method of HS09 on a larger CMIP6 ensemble yields qualitatively similar results, at least in most cases (Fig.~S9). Similarly, the internal variability estimates of the different large ensemble agree in their order of magnitude (Fig.~S10). However, there clearly is a lot of noise in the method, urging caution. 

\subsubsection{The impact of scenarios}
Many large ensembles are only run for the highest emission scenario (RCP8.5 or SSP5-8.5). This scenario has been shown to be very unlikely given current policy trajectories \citep{pielke_jr_plausible_2022}. Using a high emission scenario results, in most cases, in a larger signal, i.e. change over the next century, particularly for temperature extremes. This, in turn, causes internal and model uncertainty to increase. However, model uncertainty increases more quickly than internal variability (Figs.~S9 and S11). It is to be expected then that repeating this study using a lower emission scenario, such as SSP2-4.5, would lead to a lower total uncertainty and a larger relative contribution of internal variability. 

Scenario uncertainty itself is naturally also affected by the choice of scenarios. The four scenarios selected here and similar studies, were not chosen based on any particular likelihood attached to them. In fact, SSP1-2.6 and especially SSP5-8.5 represent the fringes of likely outcomes. A denser sampling of intermediate scenarios would, most likely, lead to a reduction in scenario uncertainty as computed in this framework. 

\subsubsection{Models as truth}
One assumption underlying all studies using models to estimate climate projection uncertainty is that the models are good enough to accurately represent the different types of variabilities. This, however, is a leap of faith, especially when it comes to extremes, where climate models have been shown to struggle \citep{gervais_how_2014}, even though some research suggests that the fractional change in extremes is still adequately captured \citep{martinez-villalobos_climate_2021}. Since all climate models are based on similar principles and suffer from the same issues, such as insufficient resolution, there is no guarantee that the spread given by the models encompasses the "true" behavior of the atmosphere. 

For internal variability, the large ensembles used in this study roughly agree (Fig.~S10). However, true variability might be lower or higher. To investigate this one could compare the variability of historical simulations against observations, something we did not do here. One problem is that for very rare events the sample size in global observations is quite small. 

For model uncertainty estimates to be correct, the truth would have to be within the envelop suggested by the models. For future projections, there is obviously no way to check whether this is true. If the true evolution was outside of the model distribution, the model uncertainty or error would be much larger than estimated here. There is some evidence that low-resolution climate models underestimate the change in extremes compared to convection-permitting models \citep{kendon_greater_2020}. However, because of the computational expense of such high-resolution models, these studies have, so far, only been done regionally. 

\subsection{Role of downscaling/post-processing}

In this study an implicit bias correction was done by using quantiles, estimated for each model separately. This is analogous to using a quantile mapping. This is certainly necessary as the quantile values vary greatly between models, highlighting the well known regional biases in climate models. Computing a single quantile value across all models, therefore, leads to a drastic increase in model uncertainty (Fig.~S5).

Most sophisticated applications of climate projections, however, go beyond a simple quantile mapping to yield trustworthy results. First, there are better suited downscaling/post-processing techniques available \citep{maraun_statistical_2018}. One example is a quantile delta mapping \citep{cannon_bias_2015}, which better preserves the signal of climate change, compared to a simple quantile mapping. We tested the effect of using this method on variance estimates of our three example locations (Fig.~S5) with varying effect on the relative contributions of $I$ and $M$. In Lagos, for example, the results differ drastically. Without further analysis though, it is difficult to explain these differences.

Next, there is the question of model performance. It is widely recognized that not all climate models perform equally well in different aspects. A key metric is climate sensitivity. Some CMIP6-class models have climate sensitivities deemed far outside the likely range as estimated by a range of methods \citep{zelinka_causes_2020}. Here, we performed a very simple adjustment by simply removing CanESM5, the "hottest" CMIP6 model, from our analysis. To narrow the spread in climate sensitivities the latest IPCC report \citep{arias2021climate} relies heavily on emulators to merge different climate models with appropriate likelihoods. A similar procedure would also affect model uncertainty, most likely decreasing it.

Lastly, as already mentioned in the previous sub-section, dynamical downscaling using convection permitting-models (e.g. \cite{tapiador_regional_2020}) could provide more realistic estimates of internal variability. While some multi-model regional downscaling experiments have been done (e.g. \cite{jacob_regional_2020}), so far and to our knowledge, no convection-permitting large ensembles exist, which would be required to partition uncertainties. 

Current uncertainty partitioning studies, including this one, lag behind the state-of-the-art of climate model downscaling/post-processing. A logical next step would be to test the impact of the aforementioned methods (and more) on uncertainty partitioning. 

\subsection{The potential impact of better climate models}

One motivation of this study was to better understand the impact of improving climate models on the usefulness of their projections. As already discussed in the previous paragraphs, there are too many caveats in the methodology to accept the results here as the final word. However, some key findings are also relatively insensitive to perturbations in the method so that there should be some confidence in at least a "ballpark" estimate. 

If we assume that better climate models will reduce model uncertainty 
, what can be said about the changes to actionability? We can use the Signal-to-Noise ratio as a rough (and definitely not perfect) proxy for how actionable a forecast is. Low SNRs (below 0.5) essentially indicate that the noise obscures any actionable signal. For extreme precipitation, this is the case for most regions throughout the first half of the century. Since the noise is overwhelmingly caused by internal variability, improving models has little impact in this framework. Even towards the latter half of the century we see noise trumping signal in many continental regions. Only in some tropical regions, especially West Africa, is there a big potential improvement from reducing model uncertainty. 

One can argue that none of the current generation climate models are able to produce realistic extreme precipitation at all, thereby implying a much larger model error. However, if the SNRs suggested here are even remotely accurate, past observations (if available for a long enough period) can provide almost as accurate a picture of future extreme occurrences as even the best possible next-generation climate models, for most regions. 

For temperature, the expected signal is a lot more obvious than for precipitation, so that even with current-generation climate models we see an actionable signal for the middle of the century. Starting around the middle of the century, model uncertainty contributes significantly to total uncertainty, suggesting that there could be actionable reductions in total uncertainty from reducing model error, especially if one focuses on a single scenario.

It is important to note that there are large regional differences, which means that the potential impact of model improvements can vary greatly from one location to another. One example are coastal regions, where our diagnostics often show large gradients, suggesting that there is significant uncertainty attached to our method.

\conclusions  
In this study, we build upon the line of work started by HS09, and more specifically the large ensemble methodology of L20, by investigating the partitioning of climate model uncertainty into internal variability, model and scenario uncertainty, specifically for extreme events. Previous studies have looked at individual examples of extremes in the context of variability but here we extend this globally. 

In the introduction, we laid out three questions which we can now attempt to answer: 1) Internal variability dominates total uncertainty in climate projections of temperature extremes in the near-term, after which model and scenario uncertainty become the main contributors. For precipitation extremes, internal variability is dominant throughout the 21st century. There are regional differences, for example a stronger role of model variability in the tropics. 2) The signal-to-noise ratio is large for temperature in the tropics in the near-term and then globally from 2040 onward. For precipitation, the expected changes are lower and the internal variability is higher, so that signal-to-noise ratios above 1 do not appear until the 2070s in most regions. 3) Improving models and thereby reducing model uncertainty has a noticeable effect on projections of temperature extremes starting in the 2030s/40s. For precipitation, the potential impact is much smaller except in some isolated tropical regions. 

There are several methodological uncertainties in the approach used here, which is why this study should be seen as a rough estimate. We also used a very general definition of extreme events to enable a global study. Our sensitivity tests suggest that the results are qualitatively insensitive to the exact definition but further work is certainly required if one is to estimate the variability for specific, high-impact extremes and use cases. Further, the use of observation data to assess the fidelity of models to represent extremes would certainly be desirable for more detailed studies. 

Because internal variability is such an important contributor to climate projection uncertainty, we believe that considering its effect should be an integral part of climate modeling going forward. Specifically, this means that single-model large ensembles should become the standard for every future climate model. Using just a single realization could yield artificial signals that are actually part of natural fluctuations. Such arguments have already been made in \cite{deser2012communication} and more recently in \cite{deser_insights_2020}. 

Finally, we would like to emphasize again the importance of easy data access through the cloud. As data volumes increase from next-generation climate modeling efforts, downloading and maintaining local copies of data will become unmanageable for many users. Cloud data democratizes access for non-academic groups (like us) and researchers from lower income countries.

\codedataavailability{The code is available at https://github.com/MackenzieBlanusa/InternalVariability. All data is available freely in the cloud.} 












\authorcontribution{MB led the data analysis and writing. CL contributed to the model post-processing and plotting. SR conceived the project.} 

\competinginterests{No competing interests.} 


\begin{acknowledgements}
We thank V Balaji, Dave Farnham, Carlos Hoyos, Nicola Maher and R Saravanan for their helpful comments on this project.
\end{acknowledgements}






\bibliographystyle{copernicus}
\bibliography{internalrefs.bib}

\end{document}


\nolinenumbers

\title{Supplement for "The role of internal variability in global climate projections of extreme events"}


\Author[1, *]{Mackenzie L.}{Blanusa}
\Author[2]{Carla J.}{L\'{o}pez-Zurita}
\Author[2, $\dag$]{Stephan}{Rasp}

\affil[1]{University of Connecticut, Avery Point Campus, 1084 Shennecossett Rd, Groton, CT, 06340}
\affil[*]{Work done during an internship at ClimateAi}
\affil[2]{ClimateAi, San Francisco}
\affil[$\dag$]{Now at Google Research}





\runningtitle{Internal variability of extremes}

\runningauthor{Blanusa et al.}

\received{}
\pubdiscuss{} 
\revised{}
\accepted{}
\published{}


\firstpage{1}

\maketitle

\setcounter{figure}{0}

\renewcommand{\thefigure}{S\arabic{figure}}

\renewcommand{\thesection}{S\arabic{section}}



\section{Sensitivity to hyper-parameters in the default extreme definition}

The extreme definition used in the main text, herein referred to as the default extreme definition, relies on estimating extreme event occurrences over an aggregation period based on a quantile threshold. There are several hyper-parameters in the default extreme definition including the quantile, coarsening the data, consecutive days, aggregation period, and post processing. Sensitivity analyses have been performed for each hyper-parameter and are discussed in the subsequent sections below. The default values for each hyper-parameter are: quantile/return period = 10 years\footnote{As described in the main text, since we are using daily values rather than yearly maximums, our extremes do not correspond to the typical definition of return periods. However, for convenience, we still use the terminology here.}, coarsening = 1 day, consecutive days = 1, aggregation period = 10 years. For each sensitivity analysis, only one parameter was altered at a time and the rest of the hyper-parameters were set to their default values, unless stated otherwise. We use the same three example locations as in the main text. The model percent contribution to the total uncertainty (neglecting scenario uncertainty), $M\%$, is the variable used to assess the sensitivity of each hyper-parameter.  
\begin{equation}
    \text{Model \% Contribution} = (M(t,l) / (M(t,l) + I(t,l))) * 100
\end{equation}


\subsection{Quantile/"Return Period"}
\begin{figure}[h!]
\includegraphics[width=\linewidth]{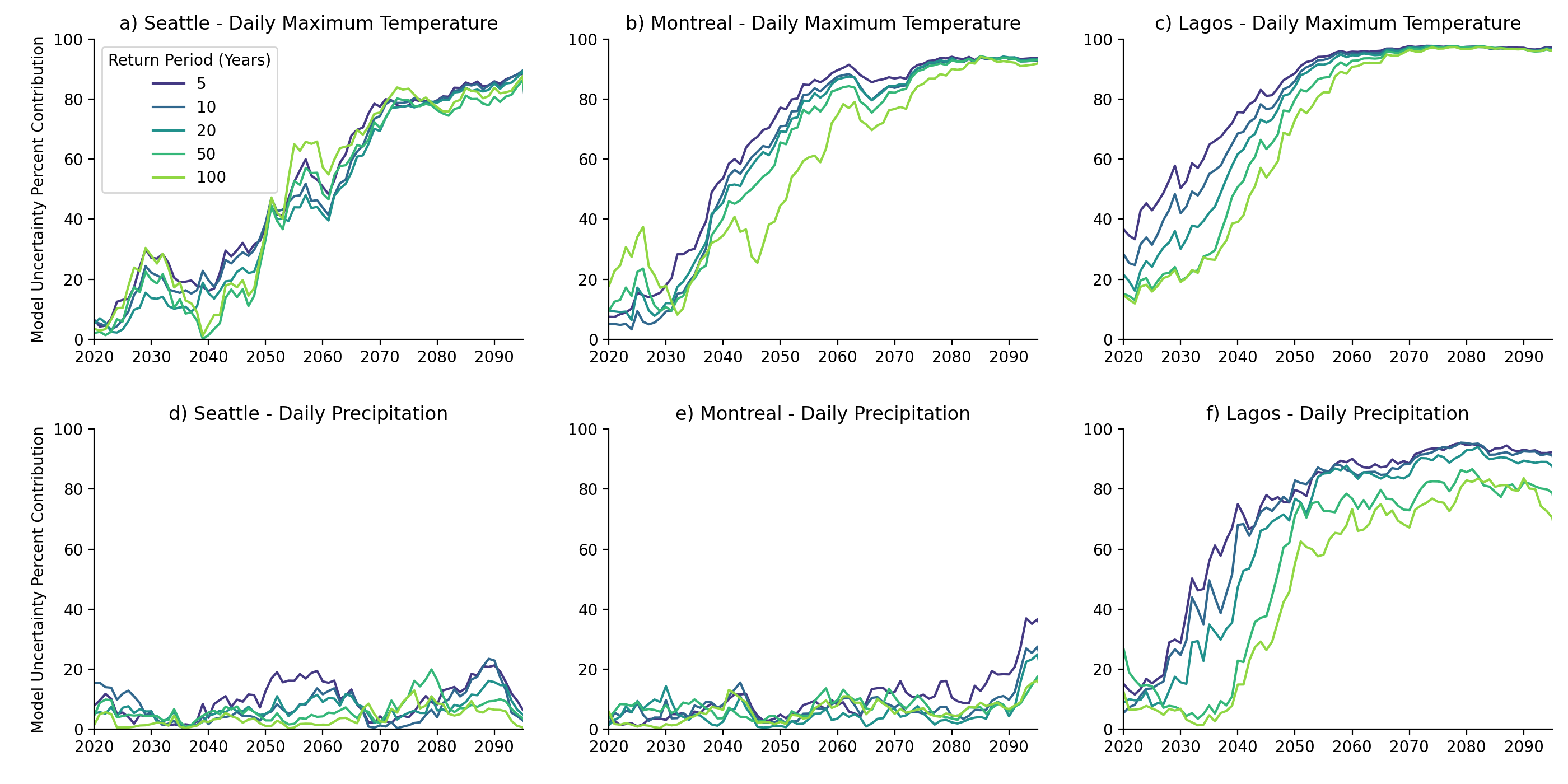}
\caption{Model \% contribution from 2020-2095 for various "return periods". Return periods of 5, 10, 20, 50, and 100 years were tested. Top row: daily maximum temperature and bottom row: daily precipitation for three locations: Seattle, U.S.A (a,d), Montreal, Canada (b,e), and Lagos, Nigeria (c,f).}
\label{fig:sensitivity return period}
\end{figure}

The "return period" was set to 5, 10, 20, 50, and 100 years while keeping all other parameters at their default values. To get from the "return period" to the quantile, we first compute the expected number of events, $E$, in the length of the reference period: 
\begin{equation}
\label{eqn:expected events}
    E = \text{\# years in reference period}\, / \,\text{return period}
\end{equation}
The quantile, $q$,  can then be computed as 
\begin{equation}
\label{eqn:quantile}
    q = 1 - E\, / \,\text{\# days in reference period }
\end{equation}
Generally, barring some noise, larger "return periods" lead to a smaller fraction of model uncertainty (Fig.~\ref{fig:sensitivity return period}). This is the result of larger internal variability attached to rarer events. Seattle and Montreal daily precipitation and Seattle maximum temperature do not show a clear pattern of the quantile effecting $M\%$ due to the already large contribution of internal variability at small return periods (e.g. Fig.~2). Overall, regions where model uncertainty dominates will be more sensitive to the quantile. While there is certainly a noticeable effect of changing the quantile, the differences are small enough to not drastically change the qualitative results.

\subsection{Days Coarsened}
\begin{figure}[h!]
\includegraphics[width=\linewidth]{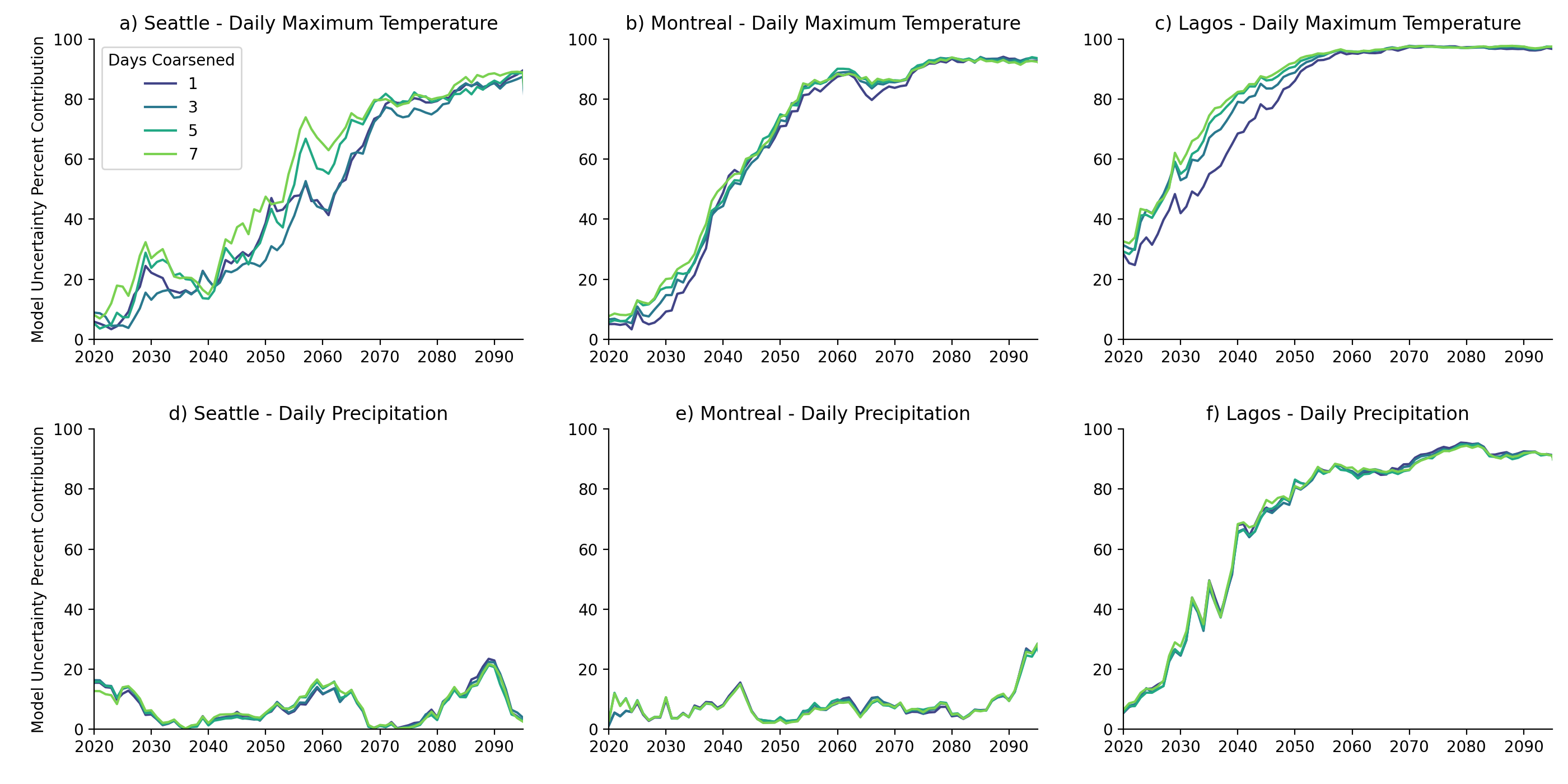}
\caption{Model \% contribution from 2020-2095 for various coarsening values. Coarsening periods of 1, 3, 5, and 7 days were tested. Top row: daily maximum temperature and bottom row: daily precipitation for three locations: Seattle, U.S.A (a,d), Montreal, Canada (b,e), and Lagos, Nigeria (c,f).}
\label{fig:sensitivity coarsen}
\end{figure}

To avoid counting consecutive days with extreme events as separate events, we can coarsen the data, so that the coarsened period is True if one or more days within indicate an extreme event. We investigate the effects of coarsening the data by setting the parameter to 1, 3, 5 and 7 days. Fig.~\ref{fig:sensitivity coarsen} clearly shows that $M\%$ is insensitive to the choice of days coarsened for both maximum temperature and precipitation. Seattle maximum temperature shows the most spread, with values converging at the end of the century.   

\subsection{Consecutive Days}
\begin{figure}[h!]
\includegraphics[width=\linewidth]{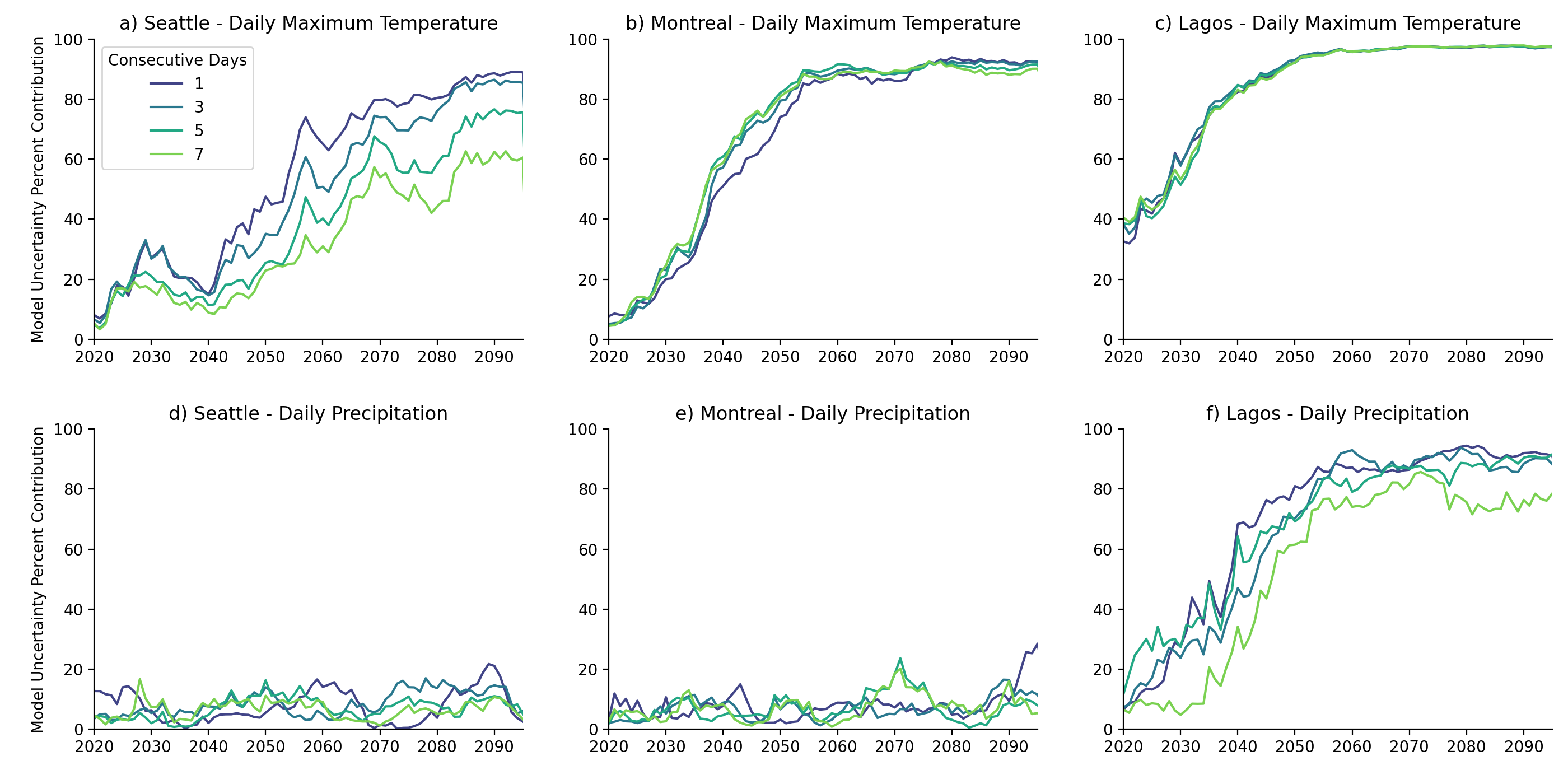}
\caption{Model \% contribution from 2020-2095 for various consecutive periods. Top row: daily maximum temperature and bottom row: daily precipitation for three locations: Seattle, U.S.A (a,d), Montreal, Canada (b,e), and Lagos, Nigeria (c,f).}
\label{fig:sensitivity consecutive}
\end{figure}

We also test the impact of using multi-day averages instead of individual daily values. This is motivated by extreme events, especially for temperature, lasting several days in many cases. Consecutive days of 1, 3, 5, and 7 were tested and all other parameters were set to their default values except the coarsening period, which was set to 7 days, since doubly counting of events is potentially a bigger issue for consecutive day extremes (co-varying both parameters ended up having a small effect in the end, however). Generally, consecutive days was found to have negligible effects on $M\%$ (Fig. \ref{fig:sensitivity consecutive}). However, there are notable effects for Seattle maximum temperature that show less $M\%$ over time for greater consecutive days. Differences for Seattle may be attributed to the larger contributions of internal variability and scenario uncertainty that persist throughout the century (Fig.~2). 

\subsection{Aggregation Period}
\begin{figure}[h!]
\includegraphics[width=\linewidth]{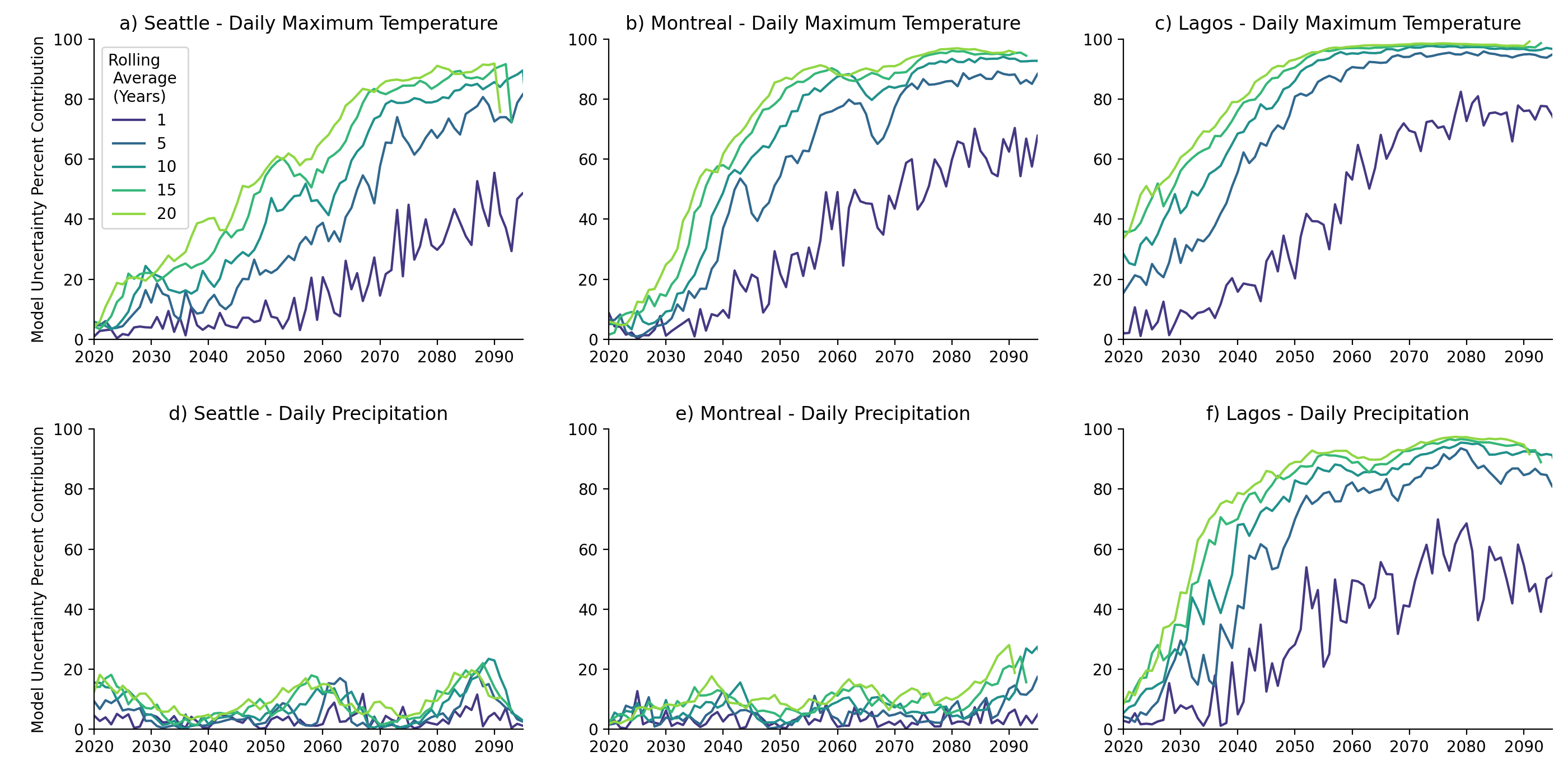}
\caption{Model \% contribution from 2020-2095 for various aggregation periods. Rolling averages of 1, 5, 10, 15, and 20 years were tested. Top row: daily maximum temperature and bottom row: daily precipitation for three locations: Seattle, U.S.A (a,d), Montreal, Canada (b,e), and Lagos, Nigeria (c,f).}
\label{fig:sensitivity rolling}
\end{figure}

We look at the average occurrence of extreme events over an aggregation period, with a default value of 10 years. As expected, this hyper-parameter has significant effects on $M\%$, where a greater aggregation period results in greater $M\%$ (Fig. \ref{fig:sensitivity rolling}). When increasing the aggregation period, the data becomes smoother, which reduces the contribution of internal variability. This sensitivity is to be expected. The default value of 10 years was selected to be consistent with previous studies \citep{hawkins_potential_2009,hawkins_potential_2011,lehner_partitioning_2020} and our own experience from delivering climate projections to customers.

\section{Sensitivity to post-processing}
\begin{figure}[h!]
\includegraphics[width=\linewidth]{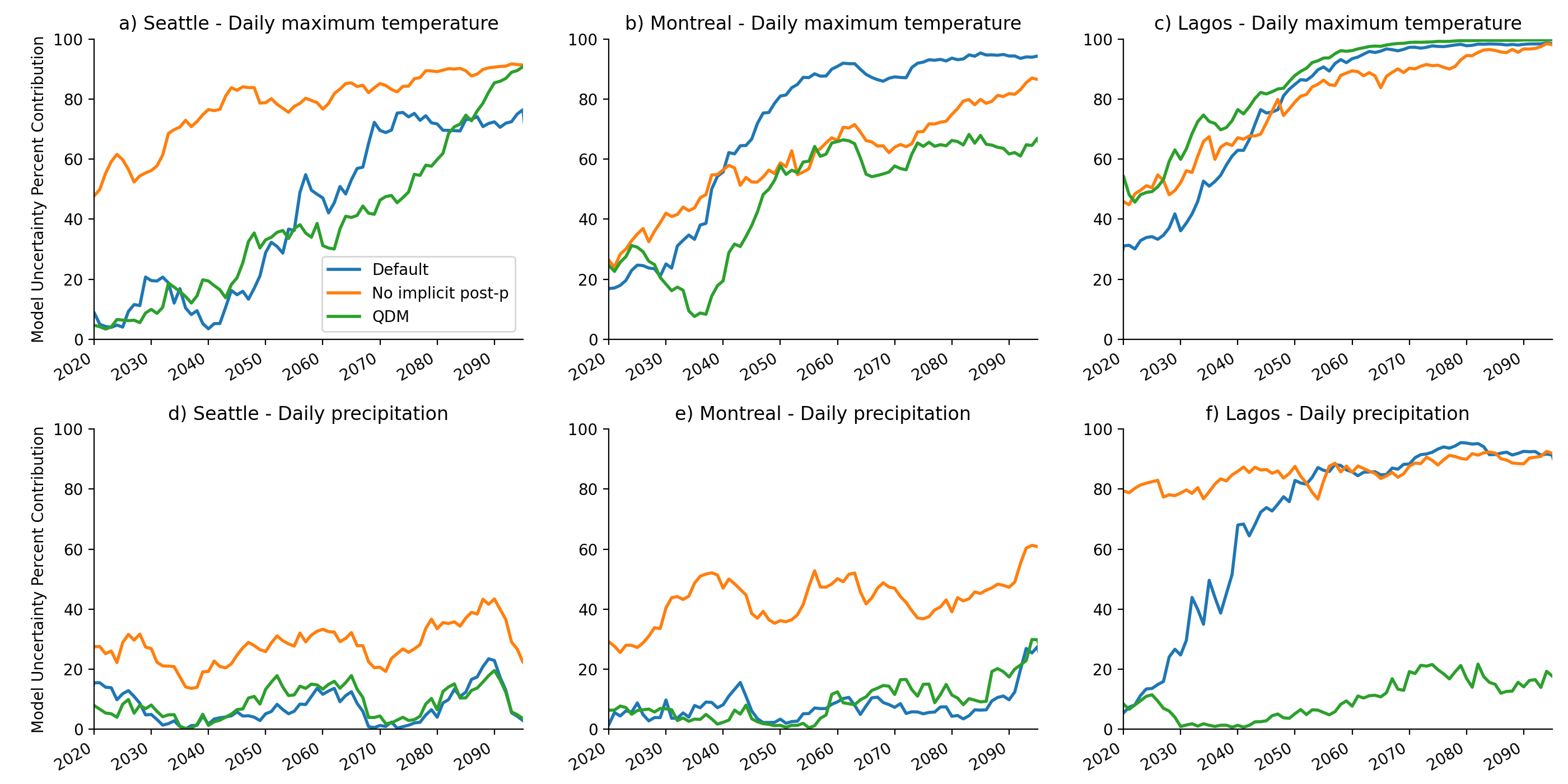}
\caption{Model \% contribution from 2020-2095 for default data (blue), default data with no implicit post-processing (orange), and post-processed data using quantile delta mapping (greeen) . Top row: daily maximum temperature and bottom row: daily precipitation for three locations: Seattle, U.S.A (a,d), Montreal, Canada (b,e), and Lagos, Nigeria (c,f).}
\label{fig:sensitivity postprocessing}
\end{figure}

Computing the quantile value for each model separately performs an implicit quantile bias correction. In Fig.~\ref{fig:sensitivity postprocessing} we compare this to taking a single quantile value across all models ("No implicit post-p"). Naturally, model variability increases drastically since climate models have large regional biases. 

We further test a state-of-the-art postprocessing method, quantile delta mapping (QDM) \citep{quantile_delta_mapping}. It was implemented taking a 30 year window as well as 3-month windows to ensure seasonality, as in the original reference. The QDM has an impact on the variance partitioning but the effect are non-systematic. This suggests that further analysis would be required to make any conclusive statements. We did not perform the QDM globally, because the algorithm is quite computationally expensive.

\section{Temperature variable}
\begin{figure}[h!]
\includegraphics[width=\linewidth]{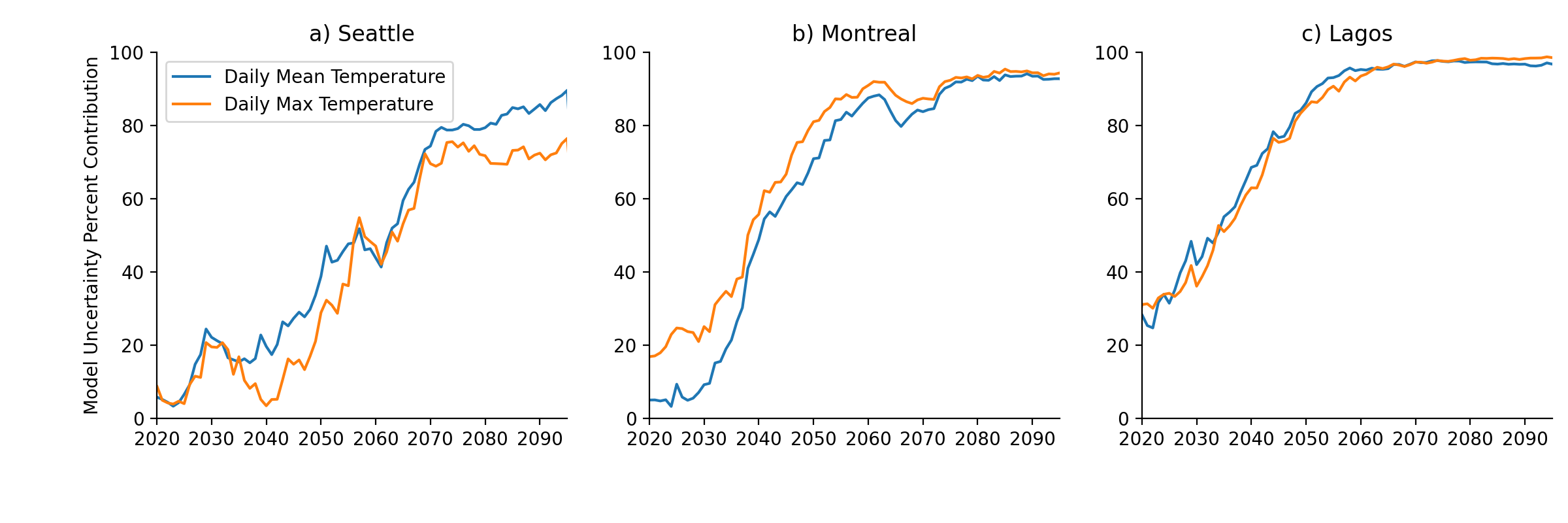}
\caption{Model \% contribution from 2020-2095 for daily mean temperature (blue) and daily maximum temperature (orange). Top row: daily maximum temperature and bottom row: daily precipitation for three locations: Seattle, U.S.A (a,d), Montreal, Canada (b,e), and Lagos, Nigeria (c,f).}
\label{fig:sensitivity tastasmax}
\end{figure}

We tested the impact of using mean temperature instead of maximum temperature on $M\%$ (Fig. \ref{fig:sensitivity tastasmax}). Overall, the differences are negligible and since the study is focused on extremes we decided to utilize daily maximum temperature.

\section{Sensitivity to extreme event definition}

We also investigated different extreme event definitions from the daily quantile definition used in the main paper. 

\subsection{T/PXx return}
First we can compute the return period based on annual maximum of maximum temperature (TXx) and daily precipitation (PXx) values rather than daily data. This is consistent with the usual definition of return periods. One problem with this is that, in a high emission scenario, this metric will quickly saturate, as all years will have a larger maximum temperature as a 1-in-10 year event in the past. Hence, the model uncertainty contribution flattens out quickly for temperature and for Lagos precipitation (Fig. \ref{fig:sensitivity extreme def}). This metric therefore is not appropriate for our context. One would have to look at much longer return periods to obtain reasonable results.

\subsection{T/PXx}

\begin{figure}[h!]
\includegraphics[width=\linewidth]{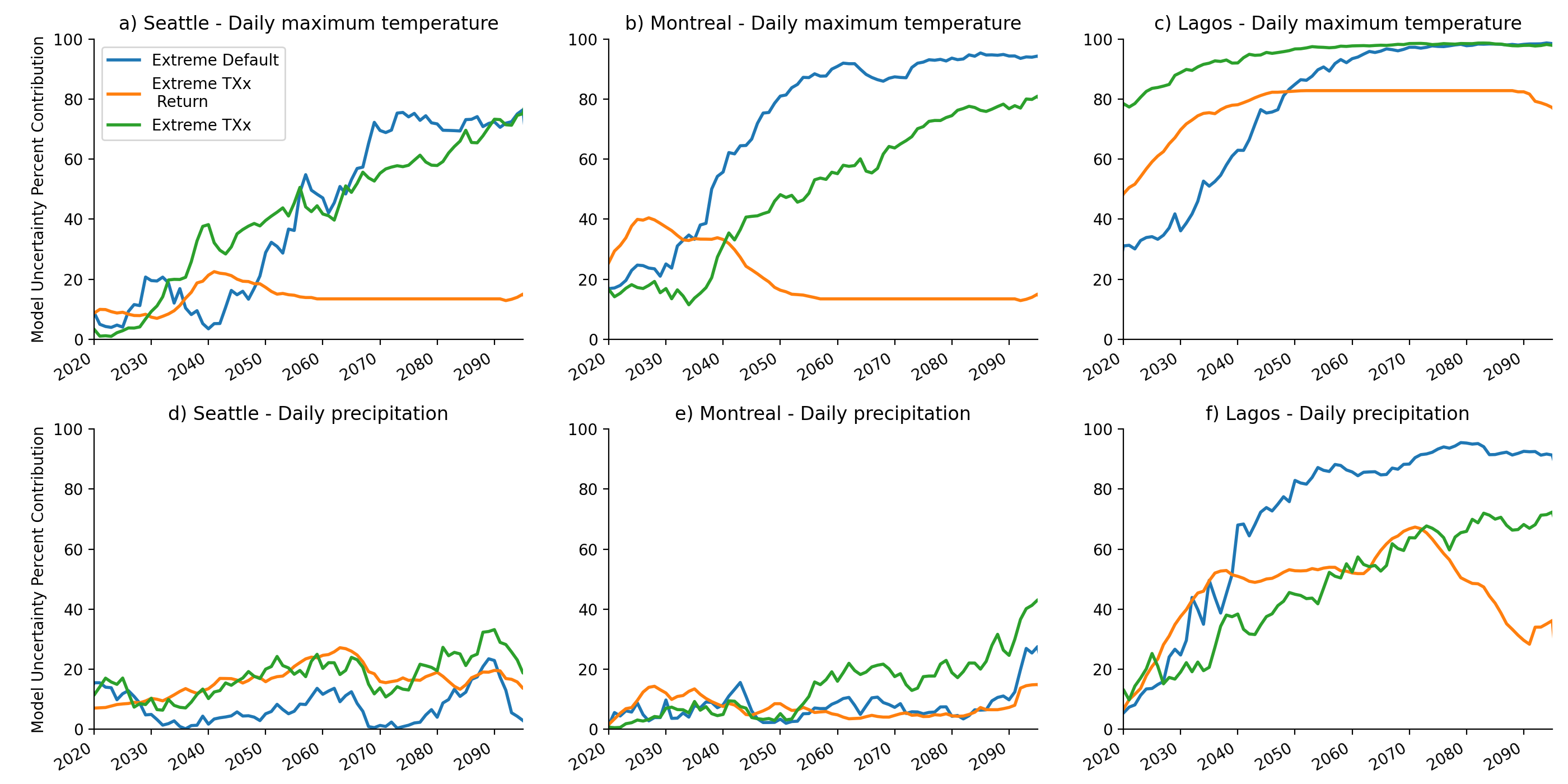}
\caption{Model \% contribution from 2020-2095 for three extreme event definitions: default (blue), extreme TXx return (orange), and extreme TXx (green). Top row: daily maximum temperature and bottom row: daily precipitation for three locations: Seattle, U.S.A (a,d), Montreal, Canada (b,e), and Lagos, Nigeria (c,f). }
\label{fig:sensitivity extreme def}
\end{figure}

One can also look at the variability of T/PXx itself, rather than the associated event occurrence. In other words, $x$ would then be the annual maximum in $^{\circ}$C and mm. In figure \ref{fig:sensitivity extreme def} we see that this has a non-negligible effect for the three example locations, going in different directions. For TXx, results stay similar in Seattle but the contribution of model uncertainty decreases in Montreal while it increases in Lagos. For PXx, the results a qualitatively unchanged for Seattle and Montreal, with a larger change in Lagos. Looking at global figures (not shown) suggests that for most locations the general picture is unchanged, however. 

\section{Sensitivity to partitioning variability method}

\begin{figure}[h!]
\includegraphics[width=\linewidth]{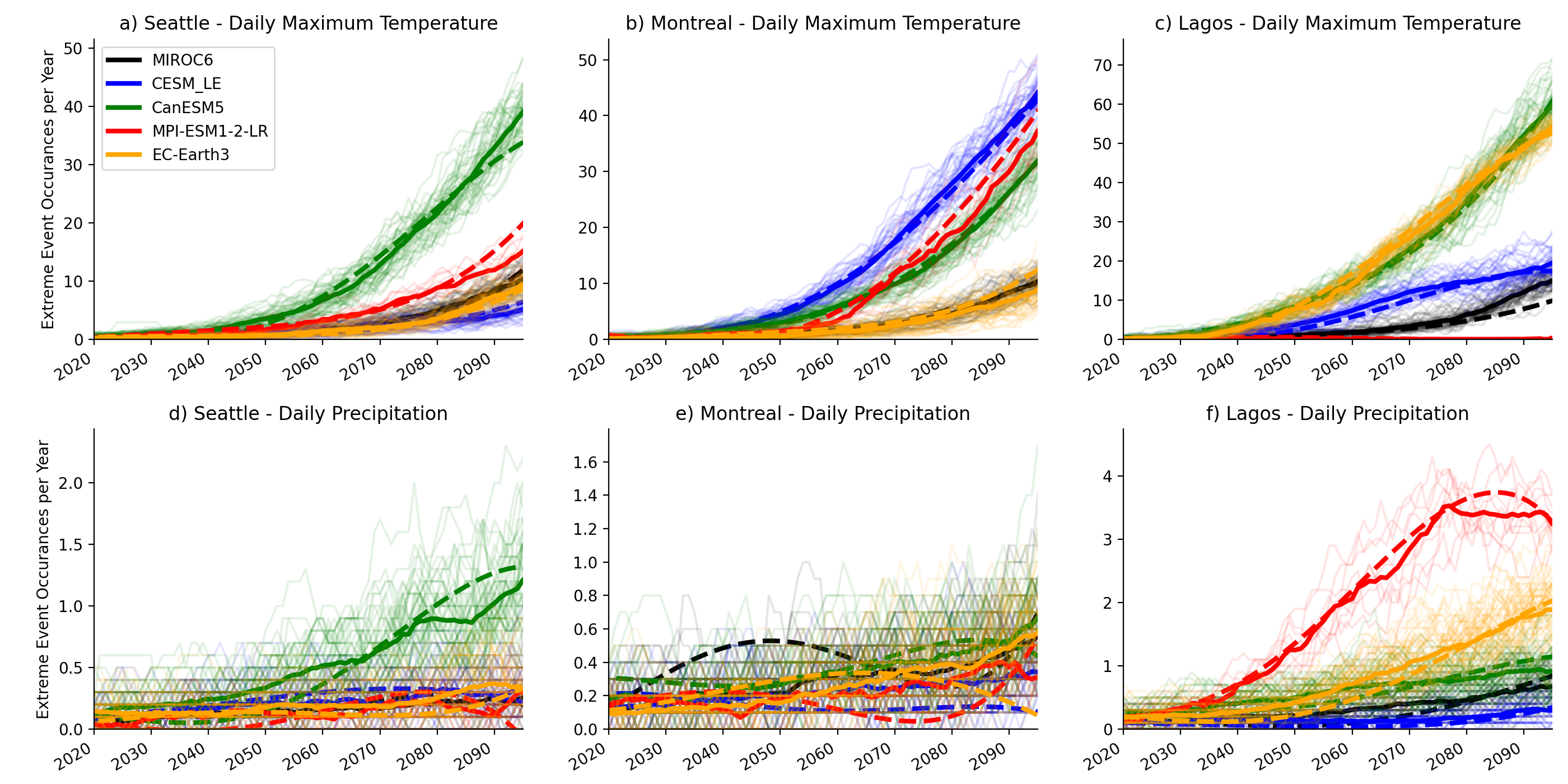}
\caption{Extreme event occurrence (days) per year from 2020-2095 as estimated from five large ensembles. Occurrences were calculated from a 10 year rolling average and divided by 10 to obtain a yearly estimate. Top row: daily maximum temperature for three locations; Seattle, U.S.A (a), Montreal, Canada (b), Lagos, Nigeria (c). Bottom row: daily precipitation for the same three regions (d-f). Dashed lines are 4th order polynomial fit for the first ensemble member of each model.}
\label{fig:ensemble fit}
\end{figure}

We can compare the large ensemble method of estimating uncertainty partitioning with the fitting method introduced by HS09. For this we follow L20 and use the first ensemble member to estimate a fourth-order polynomial fit (Fig. \ref{fig:ensemble fit}). The fit generally agrees quite well with the ensemble mean for temperature, but for precipitation the fit method picks up some noise as forced response.

\begin{figure}[h!]
\includegraphics[width=\linewidth]{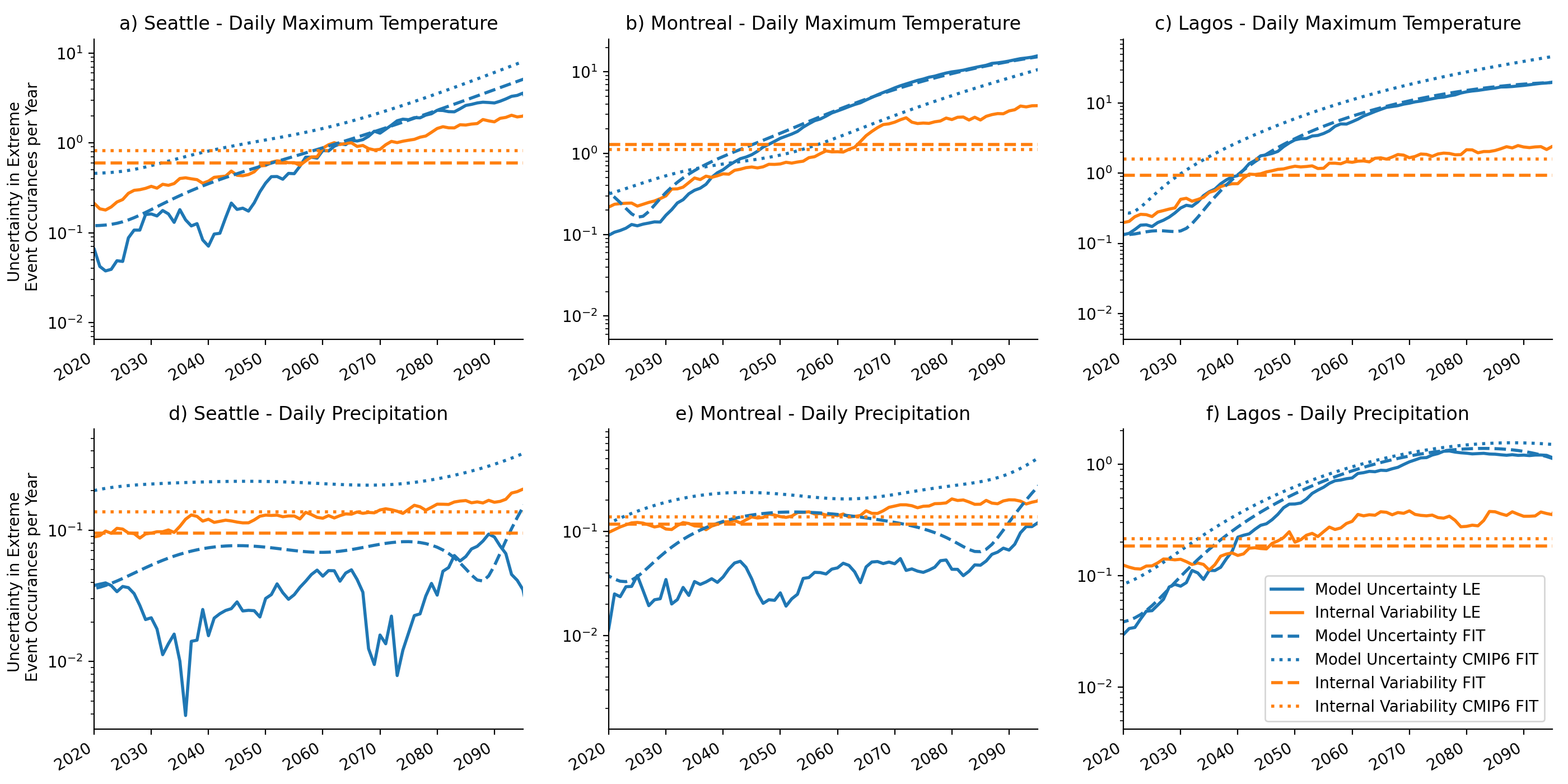}
\caption{Model uncertainty (blue) and internal variability (orange) from 2020-2095 for the large ensemble method (solid), fit method using large ensemble models (dashed), and fit method using CMIP6 single realization models (dotted). Top row: daily maximum temperature and bottom row: daily precipitation for three locations: Seattle, U.S.A (a,d), Montreal, Canada (b,e), and Lagos, Nigeria (c,f). Uncertainties were calculated with a 10 year rolling average and divided by 10 to obtain a yearly estimate.}
\label{fig:FIT vs LE}
\end{figure}

We can also apply the fitting method for our larger sample of 14 single realization CMIP6 models used for estimating the scenario uncertainty. The results for $I$ and $M$ for the different methods are shown in Figure \ref{fig:FIT vs LE}. The biggest drawback of the fitting method is that it estimates a constant $I$, which is clearly not appropriate for our extreme event definition. For temperature, $M$ estimates from the large ensemble and fitting methods generally agree in magnitude and trend. However, for precipitation large differences can be seen, a result of the fitting method picking up signal as noise. When using the fitting method on CMIP6 models larger differences can be seen, for example in Seattle temperature. We did not further investigate this behavior but also note that our quantile estimation method might be unreliable for single realization models due to low sample size (only 2 events expected in a 20 year reference period). Internal variability estimates seem to agree between the large ensemble and CMIP6 fits but are obviously not comparable to the time-changing $I$ in the large ensemble method for most locations.

\begin{figure}[h!]
\includegraphics[width=\linewidth]{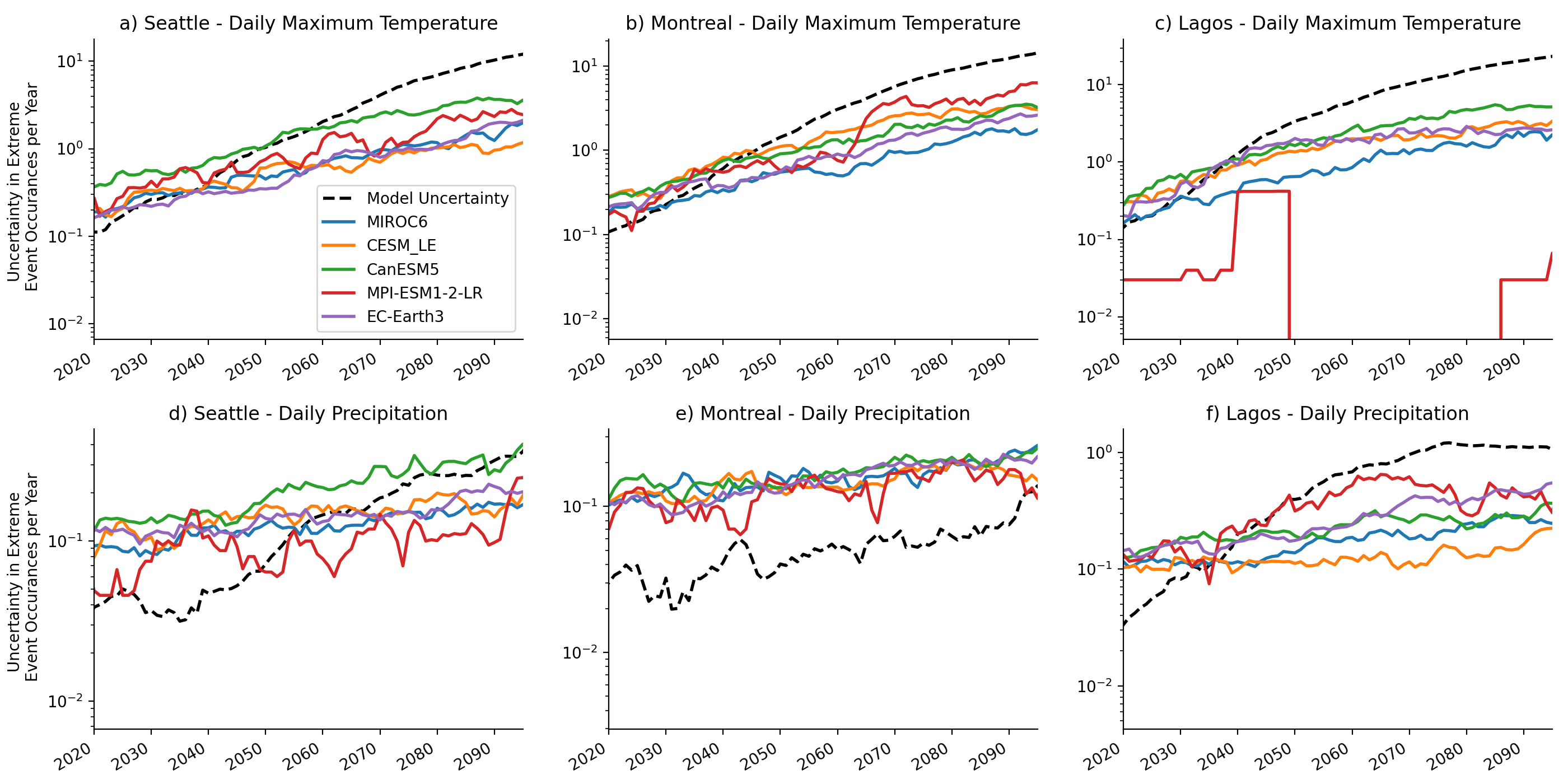}
\caption{The spread of internal variability as estimated from the five large ensembles and model uncertainty (black, dashed) from 2020-2095 for three locations: Seattle, U.S.A (a), Montreal, Canada (b), Lagos, Nigeria (c). Bottom row: daily precipitation for the same three regions (d-f). Uncertainties were calculated with a 10 year rolling average and divided by 10 to obtain a yearly estimate.}
\label{fig:LE internal}
\end{figure}

Finally, we can look at the $I$ estimates for each of the large ensembles separately (Fig. \ref{fig:LE internal}). Here we see differences up to a factor of 5 between models but also general agreement on the order of magnitude compared to $M$.

This analysis shows that in most cases the methods agree on the order of magnitude of $I$ and $M$ but also hints at large uncertainties in the method as already described in Section 4.1 in the main text.

\clearpage

\clearpage

\clearpage
\begin{figure}[h!]
\includegraphics[width=\linewidth]{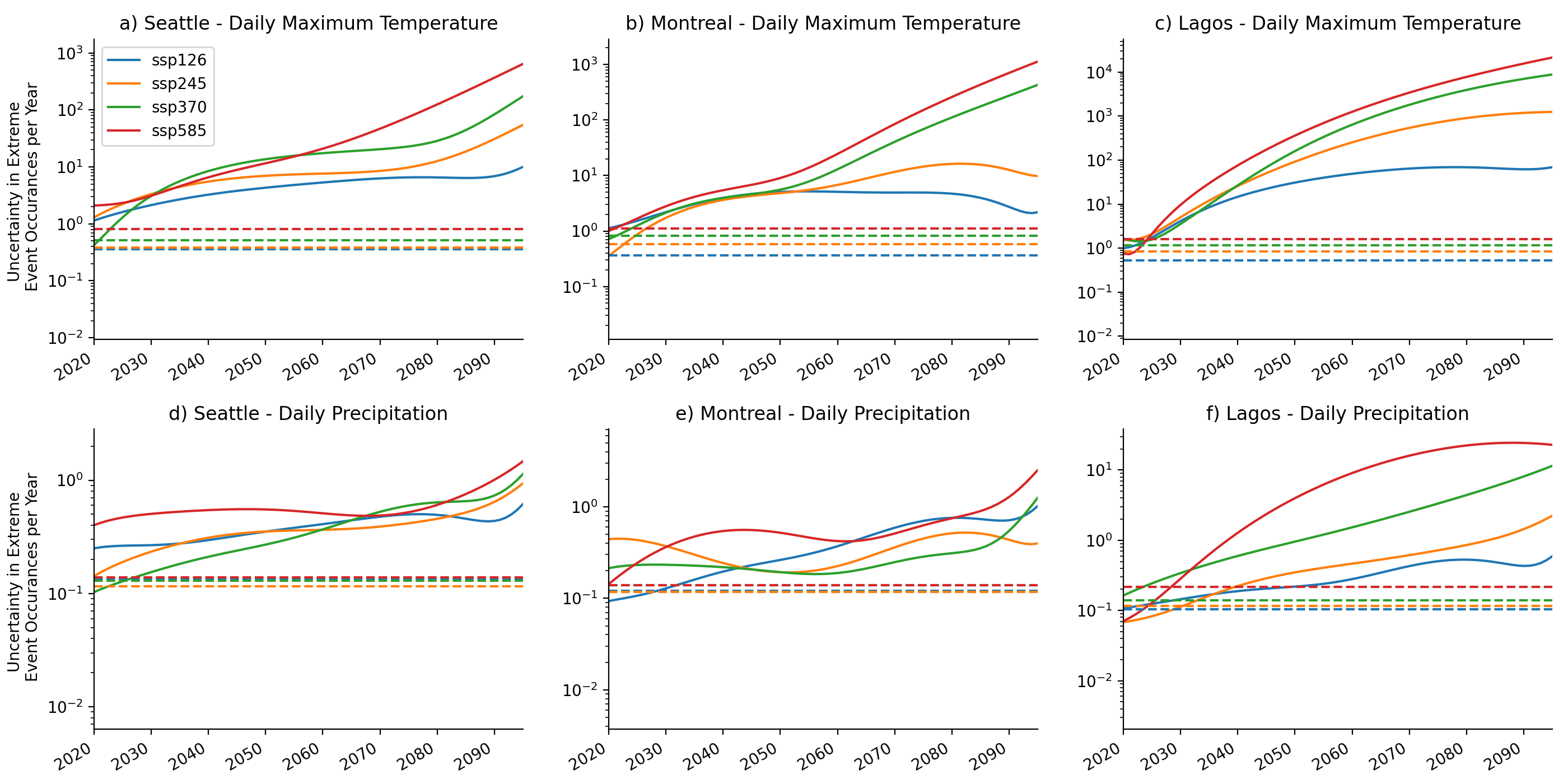}
\caption{Sensitivity analysis for the effects of scenario forcing on model uncertainty and internal variability. Each panel shows model uncertainty (solid) and internal variability (dashed) for four SSP scenarios: SSP1-2.6 (blue), SSP2-4.5 (orange), SSP3-7.0 (green), and SSP5-8.5 (red). Top row is for daily maximum temperature for three locations: Seattle, U.S.A (a), Montreal, Canada (b), and Lagos, Nigeria (c). Bottom row (d-f) is for daily precipitation for the same three locations. Uncertainties were calculated with a 10 year rolling average and divided by 10 to obtain a yearly estimate.}
\label{fig:fig14}
\end{figure}


\clearpage
\begin{figure}[h!]
\includegraphics[width=\linewidth]{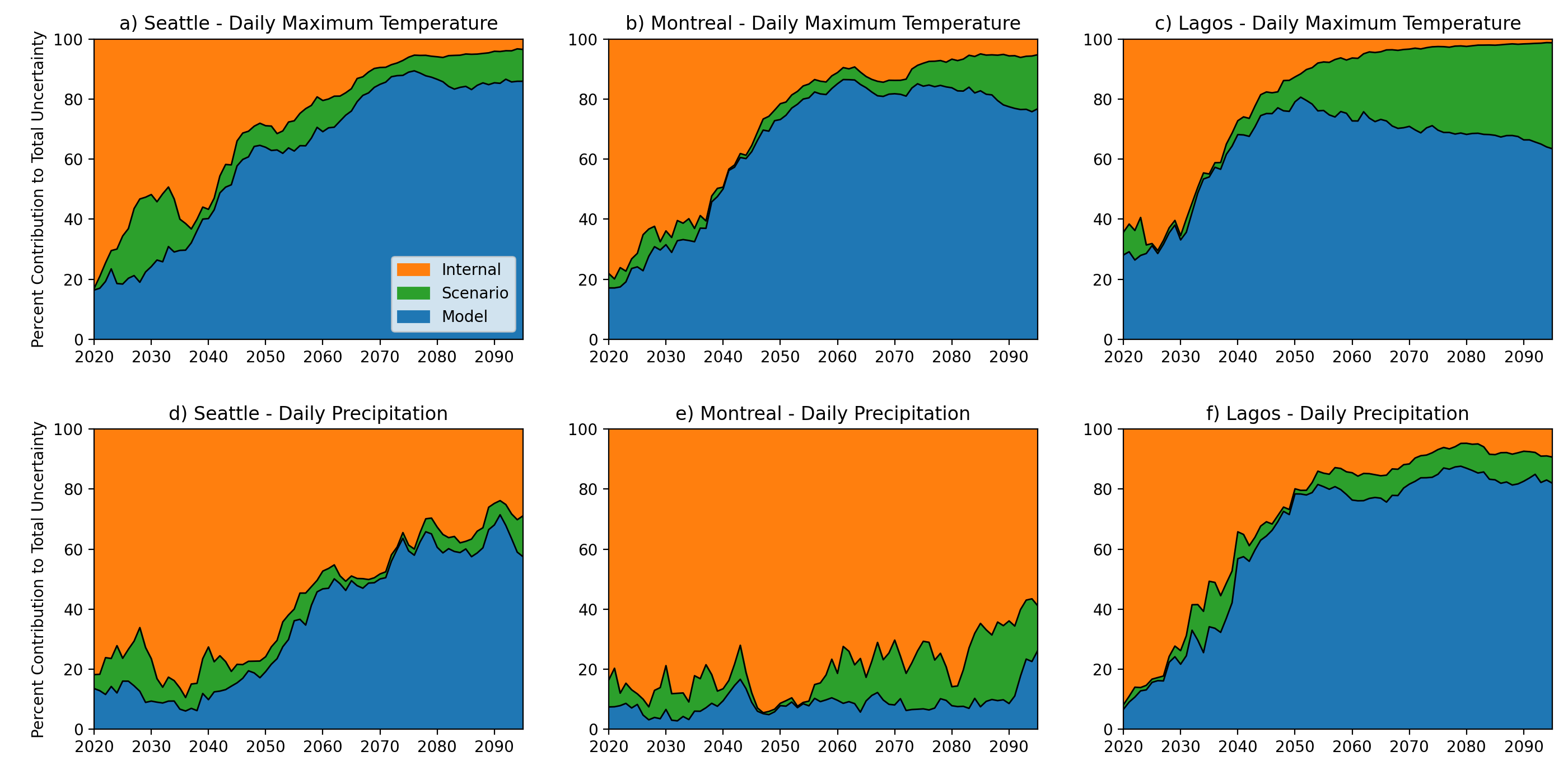}
\caption{Percent contribution to total uncertainty including CanESM5 model for model uncertainty, scenario uncertainty, and internal variability from 2020-2095 for three locations: Seattle, U.S.A (a,d), Montreal, Canada (b,e), Lagos, Nigeria (c,f). Top row (a-c) is for daily maximum temperature and bottom row (d-f) for daily precipitation. }
\label{fig:percent contribution with CanESM5}
\end{figure}

\newpage
\bibliographystyle{copernicus}
\bibliography{internalrefs.bib}